\documentclass[11pt,a4paper, twoside]{article}
\usepackage{amssymb,amsfonts,amsmath,amsthm}
\usepackage[latin1]{inputenc}
\usepackage[english]{babel}
\usepackage[left=2.7cm,right=2.7cm,bottom=2.67cm,top=2.67cm]{geometry}
\usepackage{mathtools}
\usepackage{dsfont}
\usepackage{mathrsfs}
\usepackage{natbib}
\usepackage{enumerate}
\usepackage{color}
\usepackage{setspace}
\usepackage{soul}
\usepackage[implicit=false]{hyperref}
\RequirePackage[OT1]{fontenc}
\usepackage{graphicx}
\usepackage{multirow}
\usepackage{multicol}
\usepackage{subfigure}
\usepackage{placeins}
\usepackage{rotating}
\usepackage[flushleft]{threeparttable}
\usepackage{booktabs}
\usepackage[normalem]{ulem}
\usepackage{tabularx}
\graphicspath{{figures/}}

\theoremstyle{plain}

\theoremstyle{definition}

\newcommand{\E}{\mathds{E}}
\newcommand{\F}{\mathscr{F}}
\newcommand{\R}{\mathds{R}}

\newcommand{\Z}{\mathds{Z}}
\newcommand{\eps}{\varepsilon}
\newcommand{\bs}[1]{\boldsymbol{#1}}

\newcommand{\var}{\mathrm{Var}}

\long\def\sfootnote[#1]#2{\begingroup%
\def\thefootnote{\fnsymbol{footnote}}\footnote[#1]{#2}\endgroup}

\def\bfootnote{\xdef\@thefnmark{}\@footnotetext}

\begin{document}
\pagestyle{myheadings} 
\markboth{GARTFIMA models}{Pumi, G.; Pandher, S.S. and Prass, T.}

\thispagestyle{empty}
{\centering
\Large{\bf GARTFIMA Models: A Class of Observation-Driven Models with Tempered Fractional Dynamics}\vspace{.5cm}\\
\normalsize{ {\bf Guilherme Pumi${}^{\mathrm{a,}}$\sfootnote[1]{Corresponding author. This Version: \today},\let\thefootnote\relax\footnote{${}^\mathrm{a}$Mathematics and Statistics Institute  - Universidade Federal do Rio Grande do Sul, Brazil.\\\indent ${}^b$Department of Mathematics and Statistical Sciences - University of Alberta, Canada
} Sharandeep Singh Pandher${}^\mathrm{b}$, Taiane Schaedler Prass${}^\mathrm{a}$ \\
\let\thefootnote\relax\footnote{E-mails: guilherme.pumi@ufrgs.br (Pumi); sharand1@ualberta.ca (Pandher); taiane.prass@ufrgs.br (Prass).}
\let\thefootnote\relax\footnote{ORCIDs: 0000-0002-6256-3170 (Pumi), 0000-0003-3920-2702 (Pandher), 0000-0003-3136-909X (Prass).}}\\
\vskip.3cm
}}

\begin{abstract}
This paper introduces a class of observation-driven models whose systematic component includes a tempered fractional differencing term. This specification generalizes long-range dependent models based on the fractional differencing operator, enabling a more general and robust model specification while offering theoretical advantages. We propose a partial maximum likelihood approach for parameter estimation and address hypothesis testing, confidence intervals, goodness-of-fit assessment, and both in-sample and out-of-sample forecasting. A Monte Carlo simulation study evaluates the finite-sample performance of the proposed estimation method, and an empirical application illustrates the model's practical utility.\\[.2cm]
\noindent \textbf{Keywords:} time series analysis, long-range dependence,  partial maximum likelihood, non-gaussian time series, observation-driven models.\\[.2cm]
\noindent \textbf{MSC:} 62M10, 62M20, 62F12, 62G20, 60G22.

\end{abstract}
\section{Introduction}

Time series data arising in natural and social phenomena frequently exhibit features that challenge conventional Gaussian modeling assumptions. Environmental measurements such as air pollution concentrations, streamflow levels, and energy generation are naturally constrained to positive values; relative humidity, vegetation cover indices, and mortality or morbidity rates are restricted to the unit interval $(0,1)$; while counts of adverse health events, species abundance, or extreme weather occurrences are inherently discrete. Although Gaussian linear models, such as ARMA processes, are sometimes employed for such data, their use carries tangible consequences. Perhaps the most well-known drawback is that out-of-sample forecasts can fall outside the natural bounds of the data, producing physically implausible predictions. A clear illustration is provided by \cite{grande}, who modeled influenza dynamics in Brazil using Gaussian linear regression within a peak-forecasting framework. Despite achieving good performance for peak prediction, their approach yielded negative forecasts for valleys -- a region of no interest in their study, but symptomatic of the limitations of Gaussian approximations for bounded data.

A related and particularly hazardous practice is data transformation. Transformations can drastically alter the dependence structure of a time series, especially under long-range dependence (LRD) \citep{Sangs}, and should be avoided whenever possible. Given the pitfalls of both Gaussian approximations and ad hoc transformations, it is natural to seek models that respect the data's inherent scale while preserving its dependence structure. The class of Generalized Autoregressive Moving Average (GARMA) models \citep{Benjamin2003} offers a principled solution to this problem. By embedding ARMA dynamics into a generalized linear model framework, GARMA provides a comprehensive and flexible approach that naturally accommodates bounded, double-bounded, and count outcomes, as well as covariates and asymmetry -- unlike generic non-Gaussian state-space models, which often sacrifice interpretability.

Under the GARMA framework, an observation-driven model is specified by first prescribing a conditional distribution for the time series (the random component), say $f$, depending on a measure of interest $\mu_t$ -- which may be the conditional expectation, median, quantile, or other functional -- and one or more distributional parameters. A generalized linear model (GLM) structure is then prescribed for $\mu_t$, say $g(\mu_t) = \mathbf{X}_t'\boldsymbol{\beta} + \tau_t$, where $\mathbf{X}_t$ denotes a vector of covariates and $\tau_t$ is a term responsible for controlling the dependence structure (the systematic component). When $\tau_t$ follows an ARMA-like recursion, the model reduces to the GARMA class \citep{Benjamin2003}. The GARMA family has since spawned numerous specialized models, including those for $(0,1)$-valued outcomes \citep{Rocha2009, Ribeiro2026, Pumi2026}, positive-valued data \citep{Rbayer, SantosCribariNeto2024, ptsrarxiv}, and count time series \citep{airlane, Hossain2022, CamaraAMMOD}.

Long-range dependence is a pervasive feature in many empirical time series, characterized by non-summable autocovariances and spectral density divergence at low frequencies. The most widely applied LRD models are the Gaussian ARFIMA models \citep{Granger1980, Hosking1981}, which are characterized by a hyperbolic decay of the autocorrelation function. In an ARMA setting, such decay can only be approximated by high-order models with near-unit roots. We refer the reader to \cite{palma2007} for a comprehensive treatment of LRD. Since ARFIMA models are linear, they are inherently limited to unbounded time series and cannot properly accommodate positive, double-bounded, or count data. This limitation motivated the pioneering work of \cite{PUMI2019}, and later \cite{helen} and, in a more general form, \cite{Pandher}, to extend the GARMA framework by prescribing an ARFIMA-like structure for $\tau_t$, thereby allowing the model to accommodate LRD in the systematic component. These models are generically referred to as GARFIMA models.

Despite their desirable properties and wide applicability, ARFIMA models still inherit certain drawbacks associated with LRD. The class of Autoregressive Tempered Fractionally Integrated Moving Average (ARTFIMA) models, introduced by \cite{Meerschaert2014}, generalizes ARFIMA models through exponential tempering. The resulting processes exhibit what is known as ``semi-long memory'' behavior: they mimic the slow decay of LRD models at short lags, but with a faster decay of dependence at large lags, thereby mitigating some of the limitations of ARFIMA modeling -- for instance, ARTFIMA models possess summable autocovariances and finite spectral density at the origin. The properties and applications of ARTFIMA models are studied in detail in \cite{Sabzikar2019}. However, like their ARFIMA counterparts, ARTFIMA models are defined in a linear fashion and thus face similar difficulties in modeling positive, double-bounded, and count time series.

In this paper, we fill this gap by extending the GARMA and GARFIMA frameworks to accommodate fractional tempering, effectively introducing a coherent approach for modeling non-Gaussian time series that exhibit semi-long memory, with particular emphasis on positive, double-bounded, and count-valued data. This is accomplished by prescribing an ARTFIMA-like structure for the systematic component $\tau_t$, thereby unifying the flexibility of the GARMA framework with the tempered long-memory properties of ARTFIMA models -- hereafter referred to as GARTFIMA models. Parameter estimation for GARTFIMA models is conducted via partial maximum likelihood (PMLE). We derive closed-form expressions for the score vector and the conditional information matrix, and discuss the large-sample properties of the PMLE under suitable regularity conditions. Estimation, simulation, prediction, and residual diagnostics for the GARTFIMA class have been implemented in the \texttt{BTSR} package \citep{BTSR}, available for the \texttt{R} computing environment \citep{r}. 

To assess the finite-sample behavior of the proposed estimator, we conduct an extensive Monte Carlo simulation study, evaluating the performance of the PMLE across a range of configurations, including scenarios where the tempering parameter is small and the fractional parameter varies. The practical utility of the proposed model is illustrated through two empirical applications: first, to daily maximum PM$_{2.5}$ concentrations from the Tidewater Regional Office monitoring station in Virginia, USA, a positive-valued environmental time series characterized by an extreme wildfire-driven event; and second, to squared log-returns of Amazon (AMZN) stock, a financial time series with stylized facts typical of asset returns. In both applications, we compare the proposed GARTFIMA model against a GARFIMA counterpart, assessing in-sample fit, residual diagnostics, and out-of-sample forecast performance. 

The paper is organized as follows. Section~\ref{sec:prop} presents the class of GARTFIMA models, including its random and systematic components, along with a discussion of its main theoretical properties. Section~\ref{sec:pmle} addresses parameter estimation via PMLE, including the score vector, the conditional information matrix, and asymptotic results. Section~\ref{sec:MC} presents the Monte Carlo simulation study, with results for two distinct scenarios. Sections~\ref{appl} and \ref{amzn} contain the empirical applications to pollution and financial data, respectively. Finally, Section~\ref{sec:end} concludes the paper with a summary and directions for future research.

\section{Proposed Model}\label{sec:prop}
Let $\{Y_t\}_{t\in\Z}$ be a stochastic process of interest taking values in $S\subseteq\R$ and let $\{\bs X_t\}_{t\in\Z}$ be a set of $r$-dimensional exogenous covariates to be included in the model. These can be either random or deterministic and time-dependent, or any combination of these. Let $\F_{t}$ denote the information ($\sigma$-field) available to the observer at time $t$, that is,   $\F_{t}:=\sigma\{\bs X_{t+1}, Y_{t},\bs X_{t},Y_{t-1},\cdots\}$, where, by convention, $\bs X_t$ denotes the observed values at time $t$ for deterministic covariates, and at time $t-1$ for stochastic ones. Let $g:S\rightarrow \R$ be a twice differentiable bijective link function. We propose an observation-driven class of model for which the random component is implicitly defined by assigning a probability density or mass function for $Y_t|\F_{t-1}$. We assume that the distribution of $Y_t|\F_{t-1}$ belong to the same exponential family in canonical form for all $t$, that is, we assign
\begin{equation}\label{exp}
f(y|\F_{t-1})=\exp\bigg\{\frac{y\vartheta_t-b(\vartheta_t)}{\nu}+c(y,\nu)\bigg\}I(y\in S),
\end{equation}
where $\nu$ is a dispersion parameter and $\vartheta_t$ in the canonical (or natural) parameter of the distribution. From elementary theory of exponential families, we denote the conditional mean by $\mu_t:=b'(\vartheta_t)=\E(Y_t|\F_{t-1})\in S$ and we can also obtain the conditional variance as $\var(Y_t|\F_{t-1})=\nu V(\mu_t) = \nu b''(\vartheta_t)$, where $V:S\rightarrow(0,\infty)$ is  the variance function. 

The next step on defining the proposed model, is to prescribe its systematic component. In the GARMA literature, it is usually specified by $g(\mu_t)=\bs X_t\bs\beta+\tau_t$, where $\bs\beta:=(\beta_1,\cdots,\beta_r)'$ is the vector of parameters related to the covariates and $\tau_t$ is a term that determine the serial dependence structure in the model. In this work we are interested in modeling $\tau_t$ using the ideas of general ARTFIMA models, as introduced in \cite{Meerschaert2014}. The statistical literature has extended long-range dependence models by incorporating tempering, which allows for semi-long range dependence. The ARTFIMA process is a prominent example of such models. Its primary objective is to capture long-term correlations while ensuring a summable covariance function, thereby enabling mathematically tractable analysis. This feature makes ARTFIMA particularly suitable for modeling real-world time series where long-memory behavior gradually decays over time.

The ARTFIMA model is defined by applying a tempered fractional difference operator to the observed process to obtain a stationary ARMA process. For a function $f(t)$ the tempered fractional difference operator is defined as
\[\Delta^{d,\lambda} f(t) := (1 - e^{-\lambda}L)^d f(t) = \sum_{j=0}^{\infty} \omega_{j}^{d,\lambda} f(t-j),\]
where $d \notin \Z$, $\lambda > 0$, $L$ is the backward shift operator such that $L^jf(t) = f(t-j)$, and the coefficients are given by
\begin{equation}\label{omegas}
\omega_{j}^{d,\lambda} := (-1)^j \binom{d}{j} e^{-\lambda j},\quad \text{ where }\quad \binom{d}{j} := \frac{\Gamma(1+d)}{j! \Gamma(1+d-j)}
\end{equation}
denotes the generalized binomial coefficient defined via the gamma function $\Gamma(\cdot)$. A process $\{Y_t\}_{t\in \Z}$ is said to follow an ARTFIMA$(p,d,\lambda,q)$ model if the filtered process $\Delta^{d,\lambda} Y_t$ follows an ARMA$(p,q)$ model, that is
\[\Phi(L)Y_t = (1 - e^{-\lambda}L)^{-d}\Theta(L)\eps_t,\]
where $\{\eps_t\}_{t\in\Z}$ is a white noise sequence with mean zero and variance $\sigma^2$, $\Phi(z) := 1 - \phi_1 z - \cdots - \phi_p z^p$ and $\Theta(z) := 1 + \theta_1 z + \cdots + \theta_q z^q$ are polynomials of degrees $p$ and $q$ respectively with no common zeros, and with all roots lying outside the unit circle to ensure causality and invertibility. 

The well-known class of ARFIMA$(p,d,q)$ model is a special case of the ARTFIMA$(p,d,\lambda,q)$ model when $\lambda = 0$. Unlike the ARFIMA model, which is stationary only for $-0.5 < d < 0.5$, the ARTFIMA model is stationary for any $d \notin \Z$ because tempering eliminates the unit root present in the untempered case. For integer $d > 0$, the tempered fractional difference operator $(1 - e^{-\lambda}L)^d$ reduces to a polynomial of degree $d$ with no zeros on the unit disk, so the ARTFIMA$(p,d,\lambda,q)$ model simplifies to an ARMA$(p+d,q)$ model. For integer $d < 0$, the definition is formally equivalent to an ARMA$(p,q-d)$ model. In both integer cases, the resulting models are causal and invertible under the standard ARMA conditions -- see \cite{artfima}.

For convenience, rewrite
\[\Phi(L)Y_t = (1 - e^{-\lambda}L)^{-d}\Theta(L)\eps_t = \sum_{k=0}^\infty c_k\eps_{t-k},\]
where the $c_{k}$'s depend on both, $d$ and $\lambda$, and are determined via
\begin{equation}\label{cks}
c_{k}:=  \sum_{j=0}^{\min\{k,q\}}\theta_j\omega_{j-k}^{-d,\lambda}, 
\end{equation}
with $\theta_0:=1$, and $\omega_{j}^{-d,\lambda}$ given by \eqref{omegas}.
We specify the systematic component of the proposed model by 
\begin{equation}\label{sys}
\eta_t:=g(\mu_t)= \bs X_t^\prime \bs\beta + \sum_{i=1}^p \phi_i \big[ g(Y_{t-i})-\bs X_{t-i}^\prime \bs \beta\big] + \sum_{k=1}^\infty c_kr_{t-k},
\end{equation}
where $\eta_t$ is the linear predictor, $\bs\beta:=(\beta_1, \cdots,\beta_r)^\prime$ is the parameter vector related to the covariates, $\bs\phi:=(\phi_1,\cdots,\phi_p)^\prime$, $c_k$ is given by \eqref{cks} with $\bs\theta:=(\theta_1,\cdots,\theta_q)^\prime$ are the AR and MA coefficients, respectively. The error term in \eqref{sys} is defined in a recursive fashion by setting
$r_t:=g(Y_t)-g(\mu_t)$. The proposed class of models, hereafter denoted GARTFIMA$(p,d,\lambda,q)$, could, in principle, be defined by prescribing $Y_t|\F_{t-1}$ according to \eqref{exp} together with a systematic component given by \eqref{sys}. In view of \eqref{sys}, it is clear that $\eta_t$ and $\mu_t$ are $\F_{t-1}$-measurable.

Considering the random component as a member of the canonical exponential family is convenient, but in practice this requirement may be too restrictive. The definition of the GARTFIMA model considering an arbitrary conditional density for $Y_t|\F_{t-1}$ follows the same steps presented above. In the literature of GARMA models, those for which the distribution is not a member of the exponential family in canonical form are often called GARMA-like models, to make the distinction clear - but the definition follow the same steps in both cases.

\section{Parameter estimation}\label{sec:pmle}

In this section we shall derive the partial maximum likelihood estimator (PMLE) for the parameters in the proposed GARTFIMA model.  Let $\{(Y_t,\bs X_{t}')\}^{n}_{t=1}$ be a sample from a GARTFIMA$(p,d,\lambda,q)$ model. Let us denote the $(p+q+r+3)$-dimensional parameter vector by $\bs\gamma^{\prime}=(\nu,d,\lambda,\bs\beta^{\prime},\bs\phi^{\prime},\bs\theta^{\prime})$. In what follows, for generality sake we allow $\nu$ to be estimated along with the other parameter models.  Define 
\begin{equation}
\ell_t(\bs\gamma) := \log\big(f(Y_t| \F_{t-1})\big) = \frac{Y_{t}\vartheta_t-b(\vartheta_t)}{\nu}+c(Y_{t},\nu)
\end{equation}
The partial log-likelihood function  is given by 
\begin{equation*}
  \ell({\bs\gamma}) = \sum_{t=1}^{n} \ell_t(\bs\gamma)  =  \sum_{t=1}^{n} \bigg[\frac{Y_{t}\vartheta_t-b(\vartheta_t)}{\nu}+c(Y_{t},\nu)\bigg]
\end{equation*}
The partial maximum likelihood estimator of $\bs\gamma$ is defined as
\begin{equation}\label{max}
\hat{{\bs\gamma}}:= \underset{{\bs\gamma} \in {\Omega}}{\mbox{argmax}}\big\{ \ell({\bs\gamma})\big\},
\end{equation}
Typically, \eqref{max} is obtained by solving the so-called normal equations. The partial score vector will be derived in the next section, but it is often the case that the normal equations have no explicit solution. In this case  we need to resort to numerical optimization to obtain the PMLE \eqref{max}. In what follows, equalities are to be understood as holding almost everywhere.
\subsection{Partial score vector}
To develop the partial score vector, we obtain the derivative of $\ell({\bs\gamma})$ given in \eqref{max} with respect to $\bs\gamma^{\prime}=(\nu,d,\lambda,\bs\beta^{\prime},\bs\phi^{\prime},\bs\theta^{\prime})$. For convenience, let $h$ denote the inverse of $b'$, so that $\vartheta_t=h(\mu_t)$. For $\gamma_j\neq\nu$, application of the chain rule and by elementary calculus, we obtain
\begin{equation*}
\frac{\partial \ell({\bs\gamma})}{\partial \gamma_j} = \sum_{t=1}^n \frac{\partial \ell_t(\bs\gamma)}{\partial \vartheta_t}\frac{\partial \vartheta_t}{\partial \mu_{t}}\frac{\partial\mu_{t}}{\partial \eta_t}\frac{\partial\eta_{t}}{\partial \gamma_j} =\frac1\nu\sum_{t=1}^n\frac{(Y_t-\mu_t)h'(\mu_t)}{ g'(\mu_t)}\frac{\partial\eta_{t}}{\partial \gamma_j},
\end{equation*}
Observe that, in terms of parameters $(\bs\beta^{\prime},\bs\phi^{\prime},\bs\theta^{\prime})'$, \eqref{sys} is the same as in \cite{PUMI2019} and \cite{Pandher}. Hence, for appropriate indexes $s$, denoting by $X_{ts}$ the $s$th coordinate of $\bs X_t$, we have
\begin{align*}
\frac{\partial \eta_{t}}{\partial \beta_s} &=  X_{ts} - \sum_{j=1}^{p} \phi_{j}X_{(t-j)s}-\sum_{k=1}^{\infty}c_k\frac{\partial \eta_{t-k}}{\partial \beta_s};\\
\frac{\partial \eta_{t}}{\partial \phi_{s}} &= g(Y_{t-s})-\bs X_{t-s}^{\prime} \bs\beta - \sum_{k=1}^{\infty}c_{k} \frac{\partial \eta_{t-k}}{\partial \phi_s};\\
\frac{\partial \eta_{t}}{\partial \theta_{s}} &=\sum_{k=s}^\infty \omega_{k-s}^{-d,\lambda}r_{t-k}-\sum_{k=1}^\infty c_k\frac{\partial\eta_{t-k}}{\partial\theta_s},
\end{align*}
whenever the infinite series converge.  We can also relate the coefficients $c_k$ here with the coefficients $\pi_k$ from a $\beta$ARFIMA specification, which makes derivation (and computational implementation) that much easier. We start by observing that $\binom{-d}{j}=(-1)^j\binom{j+d-1}{j}$.
With that in mind, it follows that
\[\omega_j^{-d,\lambda}=\pi_ke^{-\lambda j}, \]
where the $\pi_k$ are the coefficients of the expansion of $(1-z)^{-d}=\sum_{i=0}^\infty \pi_k z^k$ from the $\beta$ARFIMA model \citep{PUMI2019}. Hence,
\[\frac{\partial \omega_j^{-d,\lambda}}{\partial d}=e^{-\lambda j}\frac{\partial \pi_j}{\partial d}=e^{-\lambda j}\pi_{j}\big[\psi(d+j)-\psi(d)\big] =\omega_j^{-d,\lambda} \big[\psi(d+j)-\psi(d)\big], \]
where $\psi:(0,\infty)\rightarrow\R$ is the digamma function defined as $\psi(z)=d\log\big(\Gamma(z)\big)/dz$. By the reflection formula $\psi(d+j)-\psi(d) = \psi(1-d-j)-\psi(1-d)$, so that, from \eqref{cks} and the chain rule, we obtain
\begin{equation*}
\frac{\partial \eta_{t}}{\partial d}=\sum_{k=1}^{\infty}\bigg(r_{t-k} \sum_{i=0}^{\min \{k, q\}} \theta_{i} \omega_{i-k}^{-d,\lambda}[\psi(-d+1)-\psi(-d-k+i+1)]-c_{k} \frac{\partial \eta_{t-k}}{\partial d}\bigg),
\end{equation*}
provided that the infinite series is convergent. As for the derivative with respect to $\lambda$, we have $\partial \omega_j^{-d,\lambda}/\partial \lambda = -j \omega_j^{-d,\lambda}$, so that
\begin{equation*}
\frac{\partial \eta_{t}}{\partial \lambda}=\sum_{k=1}^{\infty}\bigg(r_{t-k} \sum_{i=0}^{\min \{k, q\}} (i-k)\theta_{i} \omega_{k-i}^{-d,\lambda}-c_{k} \frac{\partial \eta_{t-k}}{\partial \lambda}\bigg),
\end{equation*}
whenever the infinite series converge. Finally, the derivative with respect to $\nu$ is given by
\begin{equation*}
\frac{\partial \ell(\bs\gamma)}{\partial \nu}=\sum_{t=1}^n\bigg[\frac{b\big(h(\mu_t)\big)-Y_t h(\mu_t)}{\nu^2}+\frac{\partial c(Y_t,\nu)}{\partial \nu}\bigg].
\end{equation*}
upon writing $\bs\rho:=(\bs\beta^{\prime},d,\lambda,\bs\phi^{\prime},\bs\theta^{\prime})^{\prime}$, therefore, $\bs\gamma^{\prime}=(\nu,\bs\rho^{\prime})^{\prime}$. Define
\begin{equation*}
\bs{h_{1}}:=\bigg(\frac{\partial \ell_{1}({\bs\gamma})}{\partial \mu_{1}},\cdots,\frac{\partial \ell_{n}({\bs\gamma})}{\partial \mu_{n}} \bigg)^{\prime},\quad \text{and}\quad\bs{h_{2}}:=\bigg(\frac{\partial \ell_{1}({\bs\gamma})}{\partial \nu},\cdots,\frac{\partial \ell_{n}({\bs\gamma})}{\partial \nu} \bigg)^{\prime}.
\end{equation*}
Let $D_{\bs\rho}\in \R^{n\times(p+q+r+2)}$ denote the matrix whose  $(i,j)$th element is defined as
\begin{equation}\label{eq:score}
 [D_{\bs\rho}]_{i,j}:=\frac{\partial \eta_{i}({\bs\gamma})}{\partial \rho_{j}},   \quad\mbox{ and }\quad
T:=\mathrm{diag}\biggl\{\frac{\partial \mu_1}{\partial \eta_{t}},\cdots,\frac{\partial \mu_n}{\partial \eta_{n}}\biggr\} = \mathrm{diag}\biggl\{\frac{1}{g_1'(\mu_1)},\cdots,\frac{1}{g_1'(\mu_n)}\biggr\}.
\end{equation}
With these definitions, the score vector $U(\bs{\gamma})$ can be expressed as
\begin{equation*}
 U(\bs{\gamma}):=(U_{\nu}(\bs{\gamma})^{\prime},U_{\bs\rho}(\bs{\gamma})^{\prime})^{\prime},   \quad\text{with}\quad U_{\nu}(\bs{\gamma}):= {\bs{1}^{\prime}_n}\bs{h_{2}},\quad \mbox{and}\quad U_{\bs\rho}(\bs{\gamma}):=D^{\prime}_{\bs\rho} T \bs{h_{1}},
\end{equation*}    
 where $ \bs{1}_n:=(1,\cdots,1)^{\prime}\in \R^{n}$. The PMLE is obtained as a solution of the nonlinear system  $U(\bs{\gamma})=\bs{0}$, which usually does not admit a closed-form solution. Therefore, the PMLE must be obtained using numerical optimization techniques.
\subsection{Conditional information matrix}
In the sequel we derive the Fisher conditional information matrix, which will be useful later on deriving the asymptotic properties of the partial maximum likelihood estimator for the proposed model. Equalities involving random quantities are to be understood to hold almost surely. Let $H_t(\bs \gamma)$ be defined by
\[
H_t(\bs \gamma) := -\frac{\partial^2\ell_t(\bs\gamma)}{\partial \bs \gamma \partial \bs \gamma'},
\]
and observe that
\begin{equation*}
H(\bs \gamma) := -\frac{\partial^2\ell(\bs\gamma)}{\partial \bs \gamma \partial \bs \gamma'}  =  -\sum_{t=1}^{n}\frac{\partial^2\ell_t(\bs\gamma)}{\partial \bs \gamma \partial \bs \gamma'} = \sum_{t=1}^nH_t(\bs \gamma).
\end{equation*}
Also, observe that both, $H(\bs \gamma)$ and $\ell(\bs\gamma)$ depend on $n$, however, for simplicity and since no confusion will arise, we omit this dependence from the notation.

Let $I_n(\bs \gamma) := \E\big(H(\bs \gamma)\big)$ be the information matrix corresponding to the sample of size $n$ and $I^{(n)}(\bs \gamma)$ is the negative expectation of the hessian $H_t(\bs \gamma)$ averaged over all observations, that is,
\[
   I^{(n)}(\bs \gamma) = -\frac{1}{n}\sum_{t = 1}^n \E \bigg(\frac{\partial^2\ell_t(\bs \gamma)}{\partial \bs\gamma \partial \bs \gamma'} \bigg)=-\frac{1}{n}\E \bigg(\frac{\partial^2\ell(\bs \gamma)}{\partial \bs \gamma \partial \bs \gamma'} \bigg), \quad \mbox{and}  \quad I_n(\bs \gamma ) = nI^{(n)}(\bs \gamma).
\]
Now, observe that
\[
 I^{(n)}(\bs \gamma) = -\frac{1}{n}\sum_{t = 1}^n \E \bigg( \E\bigg(\frac{\partial^2\ell_t(\bs \gamma)}{\partial \bs\gamma \partial \bs \gamma'} \bigg| \F_{t-1}\bigg)\bigg)  = \frac{1}{n}\E(K_n(\bs \gamma)),
\]
with
\[
 K_n(\bs \gamma) : = -\sum_{t = 1}^n \E \bigg(\frac{\partial^2\ell_t(\bs \gamma)}{\partial \bs\gamma \partial \bs\gamma'} \Big| \F_{t-1}\bigg).
\]
The matrix $K_n(\bs \gamma)$ is known as the conditional information matrix corresponding to the sample of size $n$. Under some regularity conditions (see the discussion in the next section),
\begin{equation}\label{ah}
 \frac{1}{n}H(\bs \gamma) - I^{(n)}(\bs \gamma) \overset{P}{\longrightarrow} 0 \quad \mbox{and} \quad    \frac{1}{n}K_n(\bs \gamma) - I^{(n)}(\bs \gamma) \overset{P}{\longrightarrow} 0, \quad \mbox{as} \quad n\to \infty.
\end{equation}
Furthermore, $I^{(n)}(\bs \gamma) \, {\longrightarrow} \,  I(\bs \gamma)$, where
\[
I(\bs \gamma) := \lim_{n\to \infty} I^{(n)}(\bs \gamma) = \lim_{n\to \infty} -\frac{1}{n}\E \bigg(\frac{\partial^2\ell(\bs \gamma)}{\partial \bs \gamma \partial \bs \gamma'} \bigg),
\]
which is the analogous of the $I_1(\bs \gamma)$ matrix for i.i.d. samples. 

To derive $K_n(\bs\gamma)$, observe that, for $\gamma_i, \gamma_j\neq\nu$, and $i,j\in\{1,\cdots,p+q+r+3\}$, the second derivative of the log-likelihood can be written as
\begin{align*}
\frac{\partial^2\ell_t(\bs\gamma)}{\partial \gamma_i \partial \gamma_j} &= \frac{\partial}{\partial \mu_t}
\left( \frac{\partial \ell_t(\bs\gamma)}{\partial \mu_t}\frac{d \mu_t}{d \eta_t} \frac{\partial \eta_t}{\partial \gamma_j}\right)
\frac{d \mu_t}{d \eta_t} \frac{\partial \eta_t}{\partial \gamma_i} \\
&=\left[ \frac{\partial^2 \ell_t(\bs\gamma)}{\partial \mu_t^2}\frac{d \mu_t}{d \eta_t} \frac{\partial \eta_t}{\partial \gamma_j}
+ \frac{\partial \ell_t(\bs\gamma)}{\partial \mu_t}\frac{\partial}{\partial \mu_t}\left(\frac{d \mu_t}{d \eta_t} \frac{\partial \eta_t}{\partial \gamma_j} \right) \right]
\frac{d \mu_t}{d \eta_t} \frac{\partial \eta_t}{\partial \gamma_i}\,.
\end{align*}
Since $\mu_t=\E(Y_t|\F_{t-1})$ is $\F_{t-1}$-measurable, it follows that $\E\big(\partial \ell_t({\bs\gamma})/\partial \mu_t\big|\F_{t-1}\big)=0$ and the second term vanishes. Hence 
\begin{align*}
 [K_n(\bs\gamma)]_{i,j}  & =   \sum_{t=1}^{n}\bigg\{\bigg[\E\bigg(\frac{\partial^2 \ell_t(\bs\gamma)}{\partial \mu_t^2} \bigg| \F _{t-1} \bigg)
  \frac{\partial \mu_t}{\partial \eta_{t}} \frac{\partial \eta_{t}}{\partial \gamma_j} +   \E\bigg(\frac{\partial^2 \ell_t(\bs\gamma)}{\partial \mu_t\partial \nu} \bigg| \F _{t-1} \bigg)  \frac{\partial \nu}{\partial \gamma_j}\bigg] \frac{\partial \mu_t}{\partial \eta_{t}} \frac{\partial \eta_{t}}{\partial \gamma_i}\\
     & \qquad \qquad     + \bigg[\E\bigg(\frac{\partial^2 \ell_t(\bs\gamma)}{\partial\nu\partial \mu_t} \bigg| \F _{t-1} \bigg)
  \frac{\partial \mu_t}{\partial \eta_{t}} \frac{\partial \eta_{t}}{\partial \gamma_j}+  \E\bigg(\frac{\partial^2 \ell_t(\bs\gamma)}{ \partial \nu^2} \bigg| \F _{t-1} \bigg) \frac{\partial \nu}{\partial \gamma_j}\bigg] \frac{\partial \nu}{\partial \gamma_i}\bigg\},
\end{align*}
Notice that 
\begin{align*}
\frac{\partial^2 \ell_t({\bs\gamma})}{\partial \mu_t^2} = \frac1\nu\frac{\partial}{\partial \mu_t}\bigg[\frac{(Y_t-\mu_t)h'(\mu_t)}{ g'(\mu_t)}\bigg] = \frac1\nu\bigg[(Y_t-\mu_t)\frac{\partial}{\partial \mu_t}\bigg(\frac{h'(\mu_t)}{ g'(\mu_t)}\bigg)-\frac{h'(\mu_t)}{ g'(\mu_t)}\bigg],
\end{align*}
and so
\[\E\bigg(\frac{\partial^2 \ell_t(\bs\gamma)}{\partial \mu_t^2} \bigg| \F _{t-1} \bigg)=-\frac{h'(\mu_t)}{ \nu g'(\mu_t)}.\]
Upon recalling that $b'\big(h(\mu_t)\big)=\mu_t$, and from the chain rule, we obtain
\begin{equation}\label{eq:dyhbh}
\frac{\partial}{\partial\gamma_i}\Big[Y_t h(\mu_t)-b\big(h(\mu_t)\big)\Big]=\frac{\partial}{\partial\mu_t}\Big[Y_t h(\mu_t)-b\big(h(\mu_t)\big)\Big]\frac{\partial\mu_t}{\partial\gamma_i}
= (Y_t-\mu_t)h'(\mu_t)\frac{\partial\mu_t}{\partial\gamma_i}.
\end{equation}
Hence, 
\begin{align*}
\E\bigg(\frac{\partial^2 \ell_t(\bs\gamma)}{\partial \mu_t\partial\nu}\bigg|\F_{t-1}\bigg)=\E\bigg(\frac{\partial^2 \ell_t(\bs\gamma)}{\partial \nu\partial\mu_t}\bigg|\F_{t-1}\bigg) =-\frac{1}{\nu^2}\E\bigg(\frac{\partial}{\partial\mu_t}
 \bigg[Y_t h(\mu_t)-b\big(h(\mu_t)\big)\bigg]\bigg|\F_{t-1}\bigg) =0.
\end{align*}
Moving on, it is straightforward to show that
\begin{align}\label{nu}
\E\bigg(\frac{\partial^2 \ell_t(\bs\gamma)}{\partial \nu^2}\bigg|\F_{t-1}\bigg)
=\frac2{\nu^3}\Big[Y_t h(\mu_t)-b\big(h(\mu_t)\big)\Big]+\E\bigg(\frac{\partial^2 c(Y_t,\nu)}{\partial \nu^2}\bigg|\F_{t-1}\bigg).
\end{align} 
The second term in \eqref{nu} represents the contribution of the normalizing component of the exponential family to the Fisher information for the dispersion parameter. Its exact expression is distribution-specific --- in the Appendix we exemplify this calculation for the Gamma-ARTFIMA model, applied in the Monte Carlo simulation in Section~\ref{sec:MC}. The conditional Fisher information matrix for $\bs\gamma$ is then given by
\begin{equation}\label{eq:fisher}
K_n(\bs\gamma) = \left( \begin{array}{cccc}
K_{\nu,\nu}&\bs 0\\
\bs0& K_{\bs\rho,\bs\rho}
\end{array}\right),
\end{equation}
with   $K_{\nu,\nu}:= \bs 1_n' E_\nu\bs 1_n$ and $K_{\bs\rho,\bs\rho}:= D'_{\bs \rho}TE_\mu T D_{\bs \rho}$,
where $D_{\bs\rho}$, $T$ and $\bs 1_n$ are the matrices and the vector defined in \eqref{eq:score} and  $E_\mu $, $E_{\mu\nu}$ and  $E_\nu$ are diagonal matrices for which the $(t,t)$th element is given by
\begin{equation*}
[E_\nu]_{t,t} = - \E\bigg(\frac{\partial^2 \ell_t(\bs\gamma)}{ \partial \nu^2} \bigg| \F _{t-1} \bigg),\quad\mbox{and}\quad
[E_\mu ]_{t,t} = -\E\bigg(\frac{\partial^2 \ell_t(\bs\gamma)}{\partial \mu_t^2} \bigg| \F _{t-1} \bigg).
\end{equation*}
\section{Monte Carlo Simulation}\label{sec:MC}
In this section, we examine the finite-sample performance of the proposed PMLE method for estimating the parameters of the GARTFIMA model. There are many scenarios that could be considered, but our primary objective is evaluating the accuracy and stability of the parameter estimates under different levels of the long-memory and tempering parameters. 

The simulation study was conducted using \texttt{R} version 4.6.0 \citep{r}. Samples from the proposed GARTFIMA model were generated using version 1.1.0 of the \texttt{BTSR} package \citep{BTSR}. We consider two distinct sets of simulation. In the first scenario, we study the effects of fractional differencing and tempering in a Gamma-ARTFIMA$(1,d,\lambda,0)$ context, considering $d\in\{0.1,0.4,0.7\}$ and $\lambda\in\{0.05,0.1,0.2\}$. In the second scenario, we focused on higher-order fractional and stronger tempering settings with $d\in\{1.2,1.5,1.8\}$ and $\lambda\in\{0.5,0.8,1.1,1.4\}$, but in a Gamma-ARTFIMA$(0,d,\lambda,0)$ context. For each parameter configuration, time series of lengths $n\in\{300,500,1000,2000\}$ were generated, considering the logarithm as link function. A burn-in period of 500 observations was used to reduce dependence on initial conditions.

\subsection{Simulation Results: Scenario 1}

Simulation results for Scenario 1 are presented in Table~\ref{t:Scenario 1}, with true parameter values $\alpha=0.4$, $\phi=-0.35$, $\nu=16$, $d\in\{0.1,0.4,0.7\}$, and $\lambda\in\{0.05,0.1,0.2\}$. Overall, the PMLE performs well in most settings. As expected, estimation improves as the sample size $n$ increases, but many combinations already yield good results even for $n=300$.

Parameter $d$ is estimated accurately across all values of $\lambda$ and $d$, even for $n=300$. For $d=0.1$, median estimates range from $0.080$ to $0.114$; for $d=0.4$, from $0.402$ to $0.421$; for $d=0.7$, from $0.666$ to $0.723$. Some slight underestimation occurs when $d=0.1$ and $\lambda=0.2$ (e.g., $\hat d=0.080$ at $n=300$), but this corrects as $n$ grows. The tempering parameter $\lambda$ is well estimated in most cases, with notable exceptions. When $d=0.1$ and $\lambda\in\{0.05,0.2\}$, considerable bias persists even at $n=2000$: for $\lambda=0.05$, $\hat\lambda\approx0.062$ (overestimation); for $\lambda=0.2$, $\hat\lambda\approx0.176$ (underestimation). Interestingly, the intermediate combination $d=0.1$ and $\lambda=0.1$ is well estimated ($\hat\lambda\approx0.103$ at $n=2000$). For $d=0.4$ and $d=0.7$, estimates of $\lambda$ improve with $n$, though mild underestimation remains for $\lambda=0.05$ and $\lambda=0.2$.

The intercept $\alpha$ is generally well estimated, with median estimates close to the true value $0.4$ for most configurations. The only exception occurs when $d=0.7$ and $\lambda=0.05$, where $\hat\alpha=0.278$ at $n=300$, improving to $0.383$ at $n=2000$. The AR parameter $\phi$ is estimated with remarkable accuracy across all settings, with median estimates ranging from $-0.361$ to $-0.320$, always near the true value $-0.35$. The dispersion parameter $\nu$ is consistently well estimated regardless of $d$, $\lambda$, or sample size. Median estimates lie between $15.97$ and $16.22$, all very close to the true value $\nu=16$, with slight negative bias that diminishes as $n$ increases.

In summary, the PMLE performs excellently for most parameters. The main difficulties arise in estimating $\lambda$ when $d$ is small ($0.1$) and $\lambda$ is at the extremes ($0.05$ or $0.2$), and in estimating $\alpha$ under strong long-memory ($d=0.7$) combined with strong tempering ($\lambda=0.05$) at small sample sizes. All other combinations yield highly satisfactory results even for $n=300$.

\begin{table}[ht]
\centering
\tiny
\setlength{\tabcolsep}{4pt} 
\renewcommand{\arraystretch}{1.4}
\caption{Simulation results -- presented are the median estimates based on 1000 replicas with $\alpha=0.4$, $\phi=-0.35$, $\nu=16$, $\lambda\in\{0.05,0.1,0.2\}$, and $d\in\{0.1,0.4,0.7\}$ for sample sizes $n\in\{300,500,1000,2000\}$.}\label{t:Scenario 1}\vspace{0.3cm}
\begin{tabular}{c|c|ccccc||ccccc||ccccc}
\hline
\multirow{2}{*}{$\lambda$} & \multirow{2}{*}{$n$} &
     \multicolumn{5}{c||}{$d=0.1$} & \multicolumn{5}{c||}{$d=0.4$} & \multicolumn{5}{c}{$d=0.7$}\\ 
     \cline{3-17}
    & & $\hat d$ &$\hat\lambda$ & $\hat\alpha$ & $\hat\phi$ & $\hat\nu$
    & $\hat d$ &$\hat\lambda$ & $\hat\alpha$ & $\hat\phi$ & $\hat\nu$
    & $\hat d$ &$\hat\lambda$ & $\hat\alpha$ & $\hat\phi$ & $\hat\nu$
    \\ 
\hline
\multirow{4}{*}{0.05} & 300 & 0.109 & 0.071 & 0.397 & -0.353 & 16.19 & 0.421 & 0.087 & 0.384 & -0.356 & 16.21 & 0.666 & 0.065 & 0.278 & -0.320 & 16.06 \\ 
 & 500 & 0.113 & 0.068 & 0.400 & -0.357 & 16.12 & 0.410 & 0.068 & 0.390 & -0.354 & 16.14 & 0.687 & 0.061 & 0.323 & -0.336 & 15.97 \\ 
 & 1000 & 0.111 & 0.063 & 0.403 & -0.357 & 16.10 & 0.407 & 0.058 & 0.395 & -0.352 & 16.08 & 0.698 & 0.058 & 0.357 & -0.345 & 16.03 \\ 
 & 2000 & 0.107 & 0.062 & 0.402 & -0.353 & 16.04 & 0.402 & 0.055 & 0.397 & -0.351 & 16.04 & 0.700 & 0.054 & 0.383 & -0.348 & 16.04 \\ 
 \hline
\multirow{4}{*}{0.1} & 300 & 0.112 & 0.096 & 0.401 & -0.358 & 16.19 & 0.419 & 0.129 & 0.395 & -0.358 & 16.21 & 0.701 & 0.115 & 0.379 & -0.349 & 16.09 \\ 
 & 500 & 0.112 & 0.091 & 0.402 & -0.358 & 16.19 & 0.417 & 0.120 & 0.399 & -0.361 & 16.14 & 0.708 & 0.114 & 0.388 & -0.352 & 16.03 \\ 
 & 1000 & 0.114 & 0.107 & 0.404 & -0.358 & 16.08 & 0.414 & 0.117 & 0.399 & -0.355 & 16.08 & 0.707 & 0.108 & 0.395 & -0.352 & 16.04 \\ 
 & 2000 & 0.109 & 0.103 & 0.404 & -0.356 & 16.02 & 0.406 & 0.107 & 0.400 & -0.351 & 16.05 & 0.703 & 0.103 & 0.396 & -0.351 & 16.02 \\ 
 \hline
\multirow{4}{*}{0.2} & 300 & 0.080 & 0.058 & 0.397 & -0.341 & 16.22 & 0.422 & 0.228 & 0.400 & -0.356 & 16.17 & 0.722 & 0.232 & 0.387 & -0.353 & 16.21 \\ 
 & 500 & 0.089 & 0.083 & 0.400 & -0.349 & 16.14 & 0.412 & 0.221 & 0.400 & -0.354 & 16.11 & 0.723 & 0.220 & 0.392 & -0.354 & 16.07 \\ 
 & 1000 & 0.095 & 0.105 & 0.402 & -0.351 & 16.05 & 0.404 & 0.207 & 0.401 & -0.351 & 16.00 & 0.711 & 0.210 & 0.396 & -0.354 & 16.07 \\ 
 & 2000 & 0.102 & 0.176 & 0.402 & -0.354 & 16.00 & 0.404 & 0.205 & 0.402 & -0.351 & 16.02 & 0.705 & 0.207 & 0.399 & -0.352 & 16.02 \\ 
\hline
\end{tabular}
\end{table}

\FloatBarrier
\subsection{Simulation Results: Scenario 2} 

Simulation results for Scenario 2 are presented in Table~\ref{t:Scenario 2}, with true parameter values $\alpha=0.4$, $\nu=16$, $d\in\{1.2,1.5,1.8\}$, and $\lambda\in\{0.5,0.8,1.1,1.4\}$. Recall that this scenario considers a Gamma-ARTFIMA$(0,d,\lambda,0)$ model (i.e., no AR parameter), so only $d$, $\lambda$, $\alpha$, and $\nu$ are estimated. Overall, estimation remains good for most parameters, though some patterns differ from Scenario 1.

The long-memory parameter $d$ is estimated reasonably well across all configurations. For $d=1.2$, median estimates range from $1.173$ to $1.341$; for $d=1.5$, from $1.467$ to $1.741$; for $d=1.8$, from $1.808$ to $1.975$. Some overestimation is observed for $d=1.2$ and $d=1.5$, particularly when $\lambda$ is larger ($1.1$ or $1.4$) and sample size is small, but estimates improve toward the true values as $n$ increases to $2000$. For $d=1.8$, estimates are also slightly upward-biased at smaller $n$, but converge satisfactorily at $n=2000$ (e.g., $\hat d=1.808$ for $\lambda=0.5$ and $\hat d=1.847$ for $\lambda=1.1$).

The tempering parameter $\lambda$ is generally well estimated, with bias diminishing as $n$ increases. For $\lambda=0.5$, estimates approach $0.504$ at $n=2000$; for $\lambda=0.8$, estimates approach $0.808$; for $\lambda=1.1$, estimates approach $1.140$; for $\lambda=1.4$, estimates approach $1.469$. Notably, mild overestimation occurs for some combinations, such as $d=1.8$, $\lambda=1.4$, and $n=300$ ($\hat\lambda=1.469$ compared to true $1.4$), but this corrects with larger $n$. The intercept $\alpha$ is estimated with remarkable accuracy across all settings. Median estimates range from $0.377$ to $0.403$, always very close to the true value $0.4$, with no systematic bias. Similarly, the dispersion parameter $\nu$ is consistently well estimated, with median values between $15.93$ and $16.24$, all near the true value $\nu=16$ regardless of $d$, $\lambda$, or sample size.

\paragraph{Comparison with Scenario 1.}
Several differences emerge between scenarios. First, the absence of the AR parameter $\phi$ in Scenario 2 does not appear to harm estimation of the remaining parameters. Second, unlike Scenario 1 where $d=0.1$ presented estimation challenges for $\lambda$ at extreme values, Scenario 2 shows that $d$ values above $1$ are estimated with only mild bias that dissipates with larger samples. Third, $\alpha$ and $\nu$ remain extremely stable across all settings, even more so than in Scenario 1, likely because the model is simpler (no $\phi$).

\begin{table}[ht]
\centering
\scriptsize
\caption{Simulation results -- presented are the median estimates based on 1000 replicas with $\alpha=0.4$, $\nu=16$, $\lambda\in\{0.5,0.8,1.1,1.4\}$, and $d\in\{1.2,1.5,1.8\}$ for sample sizes $n\in\{300,500,1000,2000\}$.}\label{t:Scenario 2}\vspace{0.3cm}
\begin{tabular}{c|c|cccc||cccc||cccc}
\hline
\multirow{2}{*}{$\lambda$} & \multirow{2}{*}{$n$} &
     \multicolumn{4}{c||}{$d=1.2$} & \multicolumn{4}{c||}{$d=1.5$} & \multicolumn{4}{c}{$d=1.8$}\\ 
    \cline{3-14}
    & & $\hat d$ &$\hat\lambda$ & $\hat\alpha$  & $\hat\nu$
    & $\hat d$ &$\hat\lambda$ & $\hat\alpha$  & $\hat\nu$
    & $\hat d$ &$\hat\lambda$ & $\hat\alpha$  & $\hat\nu$
    \\ 
\hline 
 \multirow{4}{*}{0.5} & 300 & 1.246 & 0.534 & 0.397 & 16.09 & 1.514 & 0.521 & 0.395 & 16.06 & 1.836 & 0.519 & 0.377 & 15.95 \\ 
 & 500 & 1.240 & 0.525 & 0.397 & 16.07 & 1.523 & 0.519 & 0.397 & 16.05 & 1.841 & 0.525 & 0.383 & 15.93 \\ 
 & 1000 & 1.225 & 0.514 & 0.399 & 16.05 & 1.506 & 0.507 & 0.398 & 16.05 & 1.824 & 0.514 & 0.392 & 15.96 \\ 
 & 2000 & 1.202 & 0.504 & 0.400 & 16.01 & 1.502 & 0.502 & 0.400 & 16.03 & 1.808 & 0.505 & 0.396 & 15.98 \\ 
\hline 
 \multirow{4}{*}{0.8} & 300 & 1.254 & 0.823 & 0.399 & 16.14 & 1.627 & 0.877 & 0.400 & 16.07 & 1.892 & 0.848 & 0.396 & 16.14 \\ 
 & 500 & 1.246 & 0.830 & 0.400 & 16.14 & 1.588 & 0.850 & 0.399 & 16.08 & 1.841 & 0.828 & 0.400 & 16.09 \\ 
 & 1000 & 1.247 & 0.827 & 0.400 & 16.04 & 1.551 & 0.824 & 0.399 & 15.98 & 1.831 & 0.817 & 0.400 & 16.10 \\ 
 & 2000 & 1.220 & 0.813 & 0.400 & 15.99 & 1.510 & 0.804 & 0.400 & 15.99 & 1.808 & 0.808 & 0.401 & 16.03 \\ 
\hline 
 \multirow{4}{*}{1.1} & 300 & 1.262 & 1.119 & 0.400 & 16.17 & 1.698 & 1.213 & 0.398 & 16.17 & 1.932 & 1.179 & 0.399 & 16.12 \\ 
 & 500 & 1.293 & 1.169 & 0.401 & 16.11 & 1.741 & 1.248 & 0.398 & 16.07 & 1.921 & 1.166 & 0.399 & 16.10 \\ 
 & 1000 & 1.272 & 1.158 & 0.400 & 16.03 & 1.634 & 1.189 & 0.398 & 16.06 & 1.901 & 1.147 & 0.399 & 16.03 \\ 
 & 2000 & 1.251 & 1.140 & 0.400 & 15.99 & 1.563 & 1.141 & 0.399 & 16.04 & 1.847 & 1.129 & 0.400 & 16.01 \\ 
\hline 
 \multirow{4}{*}{1.4} & 300 & 1.173 & 1.375 & 0.402 & 16.13 & 1.467 & 1.375 & 0.403 & 16.04 & 1.930 & 1.469 & 0.400 & 16.24 \\ 
 & 500 & 1.341 & 1.468 & 0.402 & 16.13 & 1.476 & 1.386 & 0.402 & 15.99 & 1.954 & 1.501 & 0.401 & 16.14 \\ 
 & 1000 & 1.305 & 1.461 & 0.400 & 16.07 & 1.567 & 1.442 & 0.400 & 16.03 & 1.975 & 1.476 & 0.400 & 16.07 \\ 
 & 2000 & 1.281 & 1.469 & 0.400 & 16.03 & 1.560 & 1.438 & 0.400 & 16.02 & 1.835 & 1.418 & 0.400 & 16.03 \\ 
\hline
\end{tabular}
\end{table}
\paragraph{Summary.}
In summary, the PMLE performs well for the higher-order fractional and stronger tempering settings considered in Scenario 2. The main difficulties are slight overestimation of $d$ when $d$ is small within this scenario ($1.2$) or when $\lambda$ is large, particularly at small sample sizes. Nevertheless, at $n=2000$, all parameters are estimated with high accuracy. The intercept $\alpha$ and dispersion $\nu$ are essentially unbiased across all configurations.
\FloatBarrier
\subsection{Estimation Challenges}
Although ARTFIMA$(1,d,\lambda,1)$ models are identifiable, maximum likelihood estimation can be very challenging for certain combinations of $d$ and $\lambda$ due to weak identifiability of parameter in small samples. In our numerical experiments, we found that the same difficulties encountered in estimating ARTFIMA$(1,d,\lambda,1)$ models also apply to GARTFIMA$(1,d,\lambda,1)$ models. To understand why, we examine the coefficients $c_k$ in \eqref{cks}. When $p=q=1$, the coefficients reduce to
\[
c_k = \omega_k^{-d,\lambda} + \theta\,\omega_{k-1}^{-d,\lambda}, \qquad \text{with} \quad \omega_k^{-d,\lambda} \sim \frac{k^{d-1}}{\Gamma(d)} e^{-\lambda k}.
\]
From this expression, we observe that the three parameters $\lambda$, $d$, and $\theta$ jointly influence the same underlying mechanism: 
\begin{itemize}
 \setlength{\itemsep}{0pt}
  \setlength{\parsep}{0pt}
  \setlength{\topsep}{0pt}
\item $d$ controls the polynomial growth and long-memory persistence;
\item $\lambda$ governs the exponential damping (tempering);
\item $\theta$ locally modifies the initial shape of the sequence $\{c_k\}$.
\end{itemize}

A critical phenomenon occurs when $d$ exceeds $1$: the polynomial term dominates the first few lags, whereas the exponential tempering only becomes apparent at larger lags. As a result, increases in $d$ can be compensated by increases in $\lambda$, and changes in $\theta$ can adjust the initial coefficients, potentially masking differences between distinct $(d,\lambda)$ pairs. In practice, the optimizer frequently locates near-equivalent regions of the parameter space, i.e., 
\[(d, \lambda, \theta) \approx (d + \delta_1,\; \lambda + \delta_2,\; \theta + \delta_3),\]
yielding nearly identical log-likelihood values. This phenomenon produces the classical symptoms of weak identifiability, including:
\begin{itemize}
  \setlength{\itemsep}{0pt}
  \setlength{\parsep}{0pt}
  \setlength{\topsep}{0pt}
\item joint overestimation of $d$ and $\lambda$;
\item $\theta$ estimates migrating toward the boundaries $-1$ or $1$;
\item a nearly singular Hessian matrix, leading to inflated standard errors;
\item strong correlation among parameter estimators;
\item different initialization points converging to distinct solutions with similar likelihoods.
\end{itemize}
It is important to  emphasize that these difficulties are purely of numerical nature. Despite the identifiability of the GARTFIMA$(1,d,\lambda,1)$ model, empirical identification of the parameters in small samples remains challenging in this setting.
\section{Application to pollution data}\label{appl}

\begin{figure}[ht!]
\centering
\includegraphics[width=0.8\textwidth]{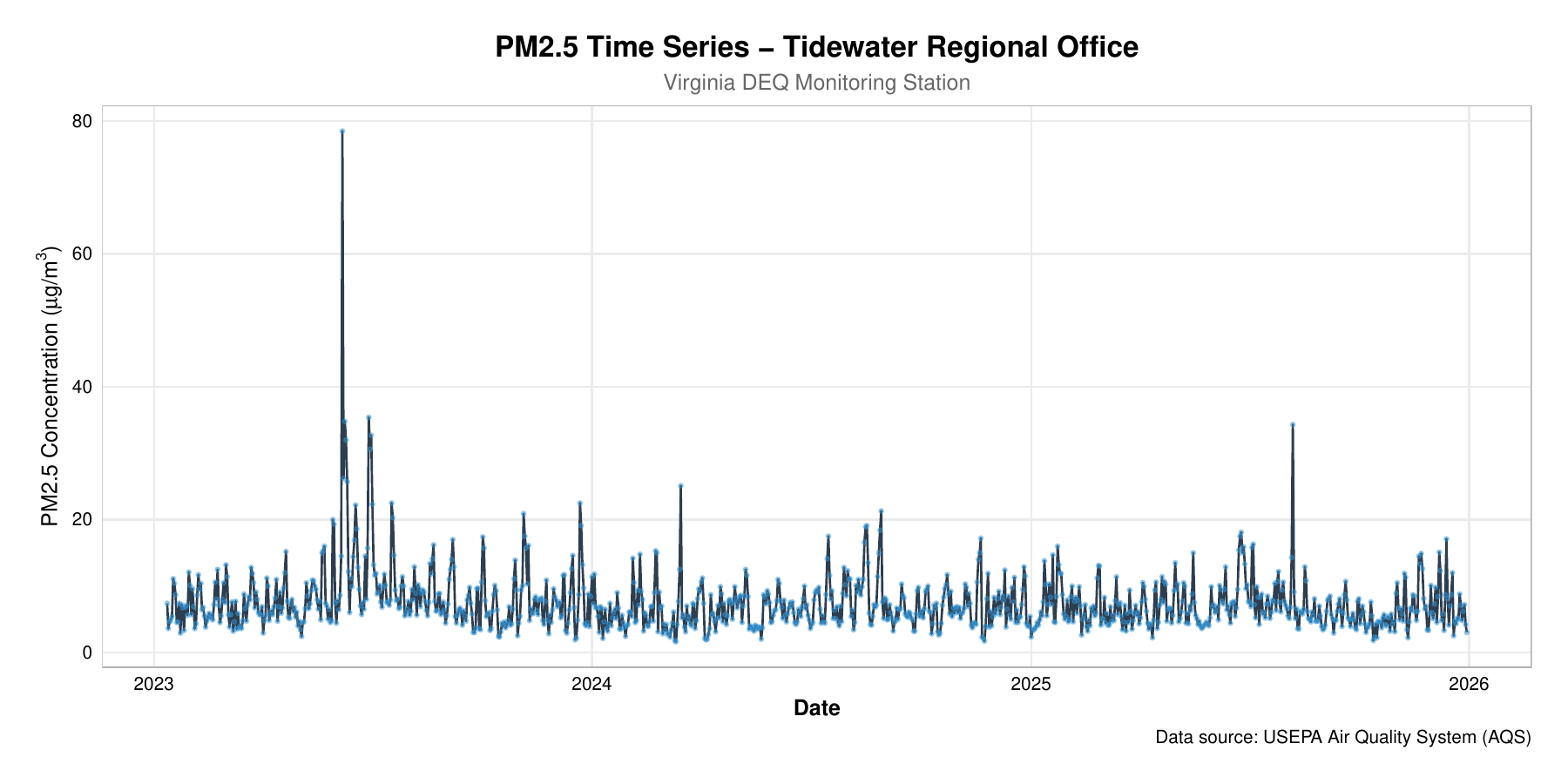}
\includegraphics[width=0.7\textwidth]{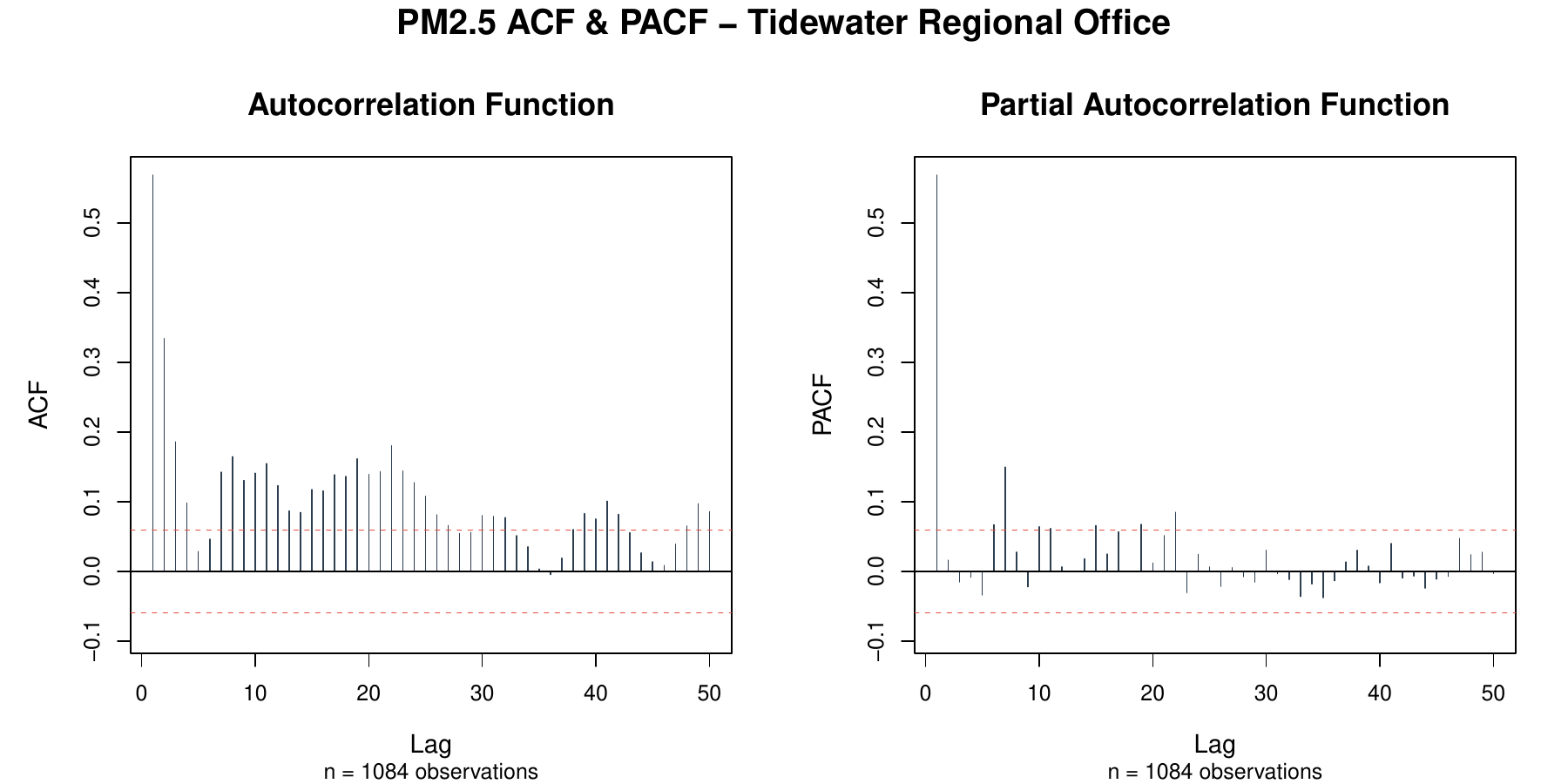}
\caption{Time series plots (top) and ACF and PACF plots (bottom).} \label{basicplots}
\end{figure}

In this section, we apply the proposed model to daily maximum (midnight-to-midnight) PM$_{2.5}$ concentrations ($\mu$g/m$^3$) from the Tidewater Regional Office monitoring station, spanning January 12, 2023, to December 30, 2025. The station, operated by the Virginia Department of Environmental Quality (DEQ), is located at 5636 Southern Blvd, Virginia Beach, within the Hampton Roads metropolitan area -- a densely populated coastal region in southeastern Virginia. The surroundings reflect a mix of urban, suburban, and marine environments. This location is subject to a complex array of pollution sources typical of dynamic urban areas. Emissions from transportation play a major role, as Hampton Roads is a key transit hub with heavy road traffic and the Port of Virginia. Industrial and commercial activities, including shipping, logistics, and manufacturing, also contribute substantially to local air pollution. These factors make the site particularly relevant for studying PM$_{2.5}$ dynamics in a multi-source setting. PM$_{2.5}$ refers to airborne particles with aerodynamic diameter $\leq$ 2.5 $\mu$m, emitted by vehicles, fires, smokestacks, and other sources. At high concentrations, these particles pose significant risks to human health. The data are publicly available from the USEPA AirNow program (\url{https://docs.airnowapi.org/}). Figure~\ref{basicplots} displays the time series, along with its ACF and PACF.

A pronounced peak in PM$_{2.5}$ concentration is visible in the time series plot, which occurred in June 7, 2023. This event is attributed to extensive wildfire smoke transported from Quebec, Canada, where record-breaking wildfires burned throughout the spring and summer of 2023~\citep{nrcan_wildfires_2023}. An upper-level low-pressure system positioned off the northeastern U.S. coast generated strong north-northwesterly winds aloft, which channeled dense smoke plumes southward into the mid-Atlantic region~\citep{noaa_smoke_2023}. This synoptic pattern, characterized by an atypical northerly flow, deviated from the region's prevailing southwesterly winds and facilitated the rapid transport of fine particulate matter over long distances. Levels of PM$_{2.5}$ attained a maximum of 78.5$\mu$g/m$^3$ at that day, a level considered unhealthy for the general population~\citep{dailypress_2023}. The intrusion of this polluted air mass resulted in hazy skies, reduced visibility, and prompted public health advisories across Hampton Roads.

The complete time series comprises 1,084 daily observations with no missing values. From these, the final 15 observations (December 16--30, 2025) were withheld for out-of-sample forecasting evaluation. The modeling sample thus consists of 1,069 observations, spanning January 12, 2023 to December 15, 2025.

Given that PM$_{2.5}$ concentrations are strictly positive, we model the data using a Gamma-ARTFIMA$(p,d,\lambda,q)$ model. For comparison, we also consider a restricted version with $\lambda = 0$, which reduces to the Gamma-ARFIMA$(p,d,q)$ model of \cite{Pandher}, but with the additional flexibility of treating the shape parameter as unknown, following the approach of \cite{PUMI2019} in the context of beta regression. All analyses were carried out in \texttt{R} 4.6.0 \citep{r}, using the \texttt{BTSR} package \citep{BTSR}. 

Model selection was performed via backward elimination based on the statistical significance of the coefficients. Starting from a full Gamma-ARTFIMA$(12,d,\lambda,12)$ specification, we sequentially removed the parameter with the highest $p$-value at each step, while retaining $\lambda$, $d$, and $\nu$ regardless of their significance. The procedure continued until all remaining coefficients were significant at the 5\% level. The same elimination strategy was applied to the Gamma-ARFIMA model. This selection process can be carried out in the \texttt{BTSR} package through the \texttt{fixed.lags} argument in the \texttt{btsr.fit} function.

The backward elimination procedure yielded a Gamma-ARTFIMA model of order $(p,q) = (4,6)$, with the restrictions $\phi_1 = \phi_2 = 0$ and $\theta_1 = \theta_3 = \theta_4 = 0$. This model will be referred to as ``Model 1'' in the following. For the Gamma-ARFIMA counterpart, the selected model was a Gamma-ARFIMA$(7,d,5)$, with $\phi_1 = \phi_2 = 0$ and $\theta_3 = \theta_4 = 0$. This will be referred as ``Model 2''. Parameter estimates and their corresponding $p$-values are reported in Table~\ref{tab:est}.

For the Gamma-ARTFIMA model, the estimate $\hat{\lambda} = 0.0305$ is statistically significant, providing empirical support for the inclusion of the tempering parameter. The estimated fractional differencing parameter is $\hat{d} = 0.6920$. Unlike in the traditional ARFIMA framework, where $d \geq 0.5$ implies non-stationarity, in the ARTFIMA model the tempering parameter $\lambda > 0$ ensures stationarity regardless of the value of $d$ \citep{artfima}. The combination of $d > 0.5$ and $\lambda > 0$ characterizes a process with tempered long-range dependence, also referred to as semi-long memory \citep{artfima}. This implies stronger persistence at shorter lags, which is subsequently attenuated at longer lags through the exponential decay induced by $\lambda$. All roots of the AR characteristic polynomial exceed 1.45 in absolute value, confirming that the absence of unit roots in the systematic component.

For the Gamma-ARFIMA model, the estimated fractional differencing parameter is $\hat{d} = 0.1253$, indicating mild long-range dependence in the systematic component. However, the AR characteristic polynomial exhibits a near unit root, with the smallest root having absolute value 1.0145. This near-nonstationary behavior, despite the modest $d$, implies that the model captures persistence primarily through the autoregressive structure rather than through fractional integration. In other words, while the series exhibits long-memory characteristics, the AR component is close to the boundary of the stationarity region, suggesting that the persistence implied through the systematic component may be better characterized by a strongly persistent AR component than by a purely fractionally integrated one. This distinction is important for interpretation, as near unit roots can lead to different implications for forecast variance and impulse response behavior compared to purely fractionally integrated processes.

A comparison of the two fitted models reveals both similarities and important distinctions. The Gamma-ARTFIMA$(4,d,\lambda,6)$ is notably more parsimonious, with 9 estimated parameters (including $\lambda$ and $\nu$), whereas the Gamma-ARFIMA$(7,d,5)$ requires 11 parameters. This difference in complexity is reflected in the model selection criteria: the Gamma-ARTFIMA yields a lower AIC (5016.8 versus 5073.3), despite having a marginally lower log-likelihood ($-2499.4$ versus $-2525.7$). These results point to the benefit of the tempering parameter $\lambda$ in capturing the persistence structure more efficiently, avoiding the need for near unit roots in the AR components. Both models share a common restricted autoregressive structure, with $\phi_1 = \phi_2 = 0$ in both specifications, and the moving average components also exhibit similarities, with $\theta_3 = \theta_4 = 0$ in both cases and both models including $\theta_2$ and $\theta_5$. Despite these structural commonalities, notable differences emerge in the estimated coefficients. For instance, $\hat{\theta}_2$ is negative ($-0.2028$) in the Gamma-ARTFIMA model but positive ($0.3439$) in the Gamma-ARFIMA model, suggesting that the interpretation of the moving average dynamics differs across the two specifications. This sign reversal likely arises from the fact that the Gamma-ARFIMA compensates for the absence of $\lambda$ by incorporating higher-order AR terms ($\phi_5$, $\phi_6$, $\phi_7$) and an additional MA term ($\theta_1$), which collectively account for the persistence that the Gamma-ARTFIMA captures more parsimoniously through the tempering mechanism. The introduction of these additional parameters may alter the correlation structure, leading to opposite signs for shared coefficients. This trade-off highlights the Gamma-ARTFIMA's ability to model long-range dependence with fewer parameters, reinforcing its advantage in terms of parsimony and interpretability.

As noted previously, the ARTFIMA and ARFIMA specifications imply markedly different autocorrelation structures. It is therefore of interest to assess which model more faithfully reproduces the dependence pattern observed in the data. However, deriving the theoretical autocorrelation function for Gamma-ARTFIMA and Gamma-ARFIMA models under a non-identity link is analytically intractable. To circumvent this difficulty, we adopt a parametric bootstrap approach to construct 95\% confidence bands for the ACF of each fitted model, and then compare these bands to the empirical ACF of the observed series.

\begin{figure}[ht!]
\centering
\includegraphics[width=0.6\textwidth]{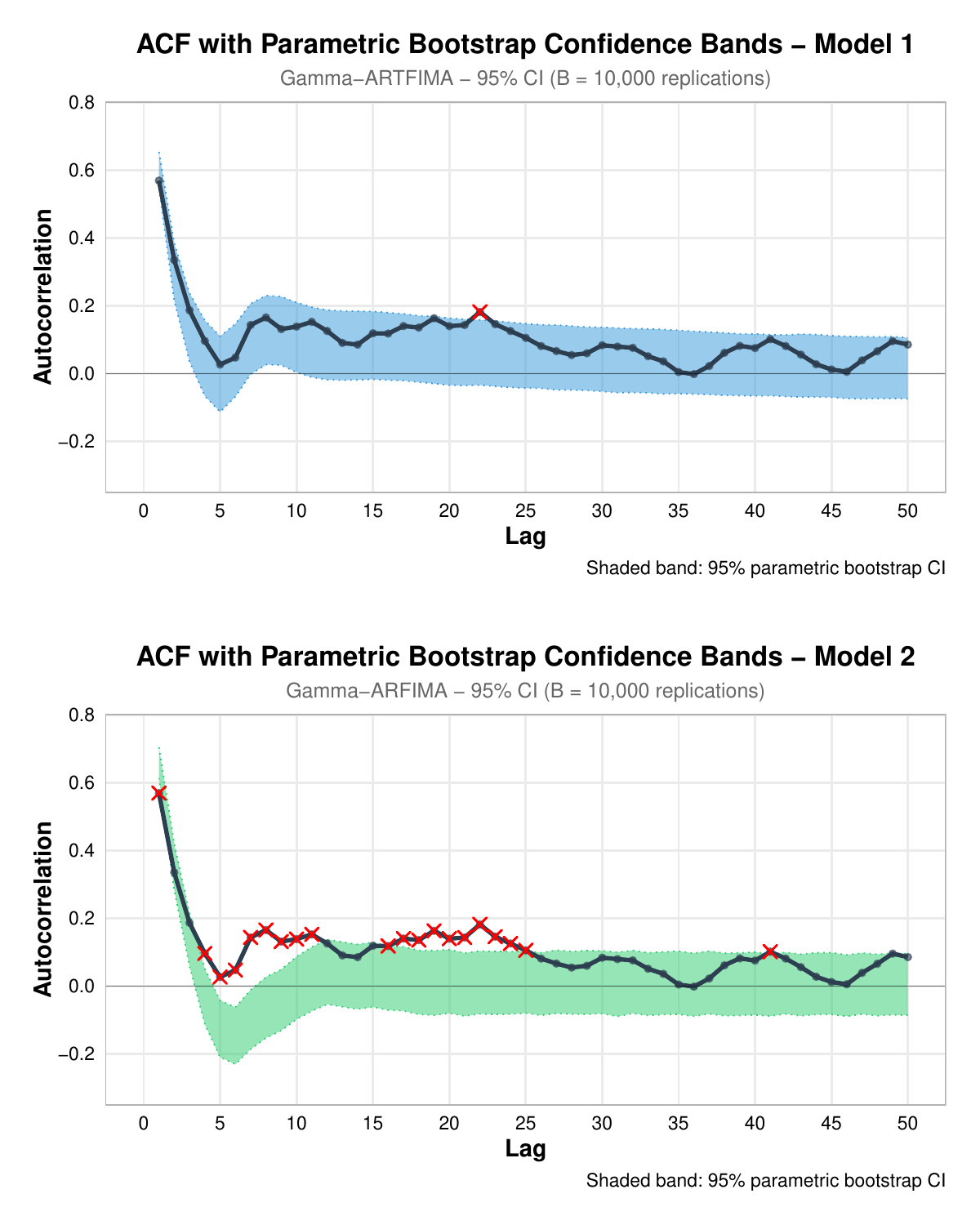}
\caption{Bootstrap confidence bands for the ACF of (top) Model 1 (Gamma-ARTFIMA) and (bottom) Model 2 (Gamma-ARFIMA) along with the observed time series ACF. Points lying outside the confidence bands are highlighted in red.} \label{confbands}
\end{figure}

The procedure is as follows. For each fitted model, we simulate a single realization of length $n = 1000$ from the estimated parameters and compute its sample autocorrelation function up to lag 50. This process is repeated $B = 10{,}000$ times, yielding a bootstrap sample of 10,000 ACF vectors, each of length 50. For each lag $k \in\{ 1, \cdots, 50\}$, the 2.5\% and 97.5\% quantiles of the 10,000 bootstrap estimates are used as the lower and upper bounds of a 95\% confidence interval. The bands thus reflect the sampling variability of the ACF under the respective fitted model, accounting for the model's estimated parameters and the inherent stochasticity of the data-generating process. Quantiles were obtained using \texttt{R}'s \texttt{quantile} function with default settings. The resulting confidence bands, along with the empirical ACF of the observed series, are displayed in Figure~\ref{confbands}. Autocorrelations falling outside the bands are highlighted in red.

The results reveal a striking contrast between the two models. For the Gamma-ARTFIMA specification, the empirical ACF lies almost entirely within the bootstrap confidence bands, with only a single lag (lag 22) falling marginally outside. This indicates that the proposed model captures the autocorrelation structure of the observed series with remarkable fidelity, reproducing both the short-term persistence and the long-term decay patterns inherent in the data. In contrast, the Gamma-ARFIMA model exhibits a substantially poorer performance: 20 out of 50 lags lie outside the confidence bands, with notable clusters at lags 1, 4--11, and 16--24. This systematic deviation suggests that the Gamma-ARFIMA model, despite its additional autoregressive and moving average parameters, is unable to adequately reproduce the dependence structure of the observed series.

\begin{table}[ht]
\centering
\caption{Estimated parameters along with respective $p-$values for the fitted GARTFIMA (Model 1) and GARFIMA models (Model 2). Also presented are the AIC and log-likelihood values.}\label{tab:est}
\vskip.2cm
\begin{tabular}{c|r|r||r|r}
\multirow{2}{*}{Parameter} &\multicolumn{2}{c||}{Model 1} &\multicolumn{2}{c}{Model 2}\\
\cline{2-5}
& Estimate & $p$-value &  Estimate & $p$-value \\
 \hline
$\alpha$   &  2.8196   &  $<0.0001$  &   2.0886  &  $<0.0001$ \\
$\phi_3$   & -0.1748   &  $<0.0001$  &   0.1740  &   0.0006 \\
$\phi_4$   & -0.1258   &  $<0.0001$  &  -0.1199  &  $<0.0001$ \\
$\phi_5$   &    ---    &     ---     &   0.3731  &  $<0.0001$ \\
$\phi_6$   &    ---    &     ---     &  -0.4659  &  $<0.0001$ \\
$\phi_7$   &    ---    &     ---     &   0.0661  &  0.0383 \\  
$\theta_1$ &    ---    &     ---     &   0.6678  &  $<0.0001$ \\
$\theta_2$ & -0.2028   &  $<0.0001$  &   0.3439  &  $<0.0001$ \\
$\theta_5$ & -0.1981   &  $<0.0001$  &  -0.5049  &  $<0.0001$ \\
$\theta_6$ & -0.1162   &     0.0003  &    ---    &   ---    \\
$d$        &  0.6920   &  $<0.0001$  &  0.1253   &  0.0070 \\ 
$\lambda$  &  0.0305   &     0.0430  &    ---    &   ---    \\
$\nu$      &  7.3068   &  $<0.0001$  &   5.8197  &  $<0.0001$ \\
\hline
AIC        &  \multicolumn{2}{c||}{5016.8} & \multicolumn{2}{c}{5073.3}\\
log-like   & \multicolumn{2}{c||}{-2499.4} &  \multicolumn{2}{c}{-2525.7}\\
\hline
\end{tabular}
\end{table}

The superior performance of the Gamma-ARTFIMA model can be attributed to the role of the tempering parameter $\lambda$. In the ARTFIMA framework, $\lambda$ controls the rate at which the long-memory effect is gradually attenuated at longer lags, effectively tempering the hyperbolic decay characteristic of purely fractionally integrated processes. This added flexibility allows the model to accommodate both the strong persistence at short lags and the faster decay at longer lags, striking a balance that the standard ARFIMA model -- with its single long-memory parameter $d$ -- cannot achieve. Consequently, the Gamma-ARFIMA model is forced to compensate for the absence of tempering by introducing higher-order AR and MA terms, which, as the bootstrap results indicate, still fail to capture the full complexity of the dependence structure. This trade-off highlights a fundamental advantage of the tempering mechanism: it enables a more parsimonious and accurate representation of the autocorrelation structure, particularly in settings where the data exhibit a mixture of short- and long-range dependence with a gradual transition between regimes. The bootstrap analysis thus provides compelling evidence that the proposed Gamma-ARTFIMA model offers a superior fit, not only in terms of standard information criteria, but also in its ability to reproduce the empirical dependence patterns of the time series.

\subsection{Residual Analysis}\label{sec:res}
For residual analysis, we consider the simple residuals defined as $r_t = y_t - \mu_t$, where $\mu_t$ denotes the conditional mean of the process. Under correct model specification, $r_t$ should behave (asymptotically) as a martingale difference sequence with respect to the natural filtration $\F_t$. To obtain the residuals, we reconstruct the conditional mean $\mu_t$ by iterating the system in \eqref{sys} using the fitted coefficients of each model.

As expected, given the nature of the models and the extreme PM$_{2.5}$ spike observed on June 7, 2023, both models substantially underestimated the peak concentration, producing a large outlier in the residual series of each model. To mitigate the influence of this single observation on the diagnostic procedures, we removed it from the residual analysis.

The residual distributions for both models appear reasonably symmetric and centered near zero. 
One can notice the presence of a few outliers in both tails, particularly in the right tail, which can be attributed to the atypical June 2023 event. Despite these departures, the overall distribution appears well-behaved.


To formally assess whether the residuals retain any remaining dependence structure, we tested the martingale difference hypothesis using three complementary approaches. First, we employed the Dominguez-Lobato martingale difference test \citep{DL}, considering both the Cram\'er-von Mises (CM) and Kolmogorov-Smirnov (KS) test statistics, as implemented in the \texttt{BTSR} package. Second, we applied the Automatic Variance Test of \cite{Choi}, using the Normal, two-point Mammen's, and Rademacher's two-point distributions to perform the wild bootstrap, as implemented in the \texttt{vrtest} package \citep{vrtest}. For all tests, $p$-values were computed based on $5{,}000$ bootstrap iterations. The resulting $p$-values are reported in Table~\ref{tab:mart}. In all cases, the tests fail to reject the null hypothesis of a martingale difference sequence, providing strong evidence that the residuals from both models are free from serial dependence and that the conditional mean dynamics have been adequately captured. This finding supports the adequacy of both specifications.


\begin{table}[ht]
\centering
\caption{$p$-values of the martingale difference tests applied to the simple residuals of the Gamma-ARTFIMA and Gamma-ARFIMA models.}\label{tab:mart}
\vskip.2cm
\begin{tabular}{c|ccccc}
\hline
 & Cp & KS & Normal  &  Mammen's  &  Rademacher's \\
\hline
Model 1 & 0.2740 & 0.1940 & 0.3704 & 0.3652 & 0.3938 \\
Model 2 & 0.0766 & 0.0912 & 0.2262 & 0.2370 & 0.2606 \\
\hline
\end{tabular}
\end{table}
\subsection{In-Sample Forecast Accuracy}

To evaluate the in-sample predictive performance of the fitted models, we consider the reconstructed conditional mean $\mu_t$ as described in Section~\ref{sec:res}. Three accuracy measures are employed to compare the models: the Root Mean Squared Error (RMSE), the Mean Absolute Percentage Error (MAPE), and the Mean Directional Accuracy (MDA). The MDA measures the proportion of times the model correctly predicts the direction of change (i.e., whether the series increases or decreases from one period to the next). Formally, it is defined as $\text{MDA} = \frac1n \sum_{t=2}^n I\big(\operatorname{sign}(y_t - y_{t-1}) = \operatorname{sign}(\hat{\mu}_t - y_{t-1})\big)$. An MDA value above 0.5 indicates that the model correctly predicts the directional movement more often than a random guess, with values closer to 1 reflecting superior directional accuracy. This measure is particularly relevant in applications where the sign of the change is of greater interest than the magnitude of the error, such as in environmental monitoring and early warning systems.

Model 1 shows marginally lower RMSE (3.6812 vs 3.7044) and MAPE (0.3372 vs 0.3486) than Model 2, indicating a slightly better fit in terms of error magnitude. Both models achieve MDA values above 0.64, correctly predicting the direction of change in approximately 65\% of the observations, with a negligible advantage for the Gamma-ARFIMA (0.6511 vs 0.6464). Overall, the differences are small, but the consistently lower error measures for the Gamma-ARTFIMA reinforce its superior fit, in line with the information criteria and residual analysis.
%

\subsection{Out-of-Sample Forecast}
\begin{figure}[ht!]
\centering
\includegraphics[width=0.8\textwidth]{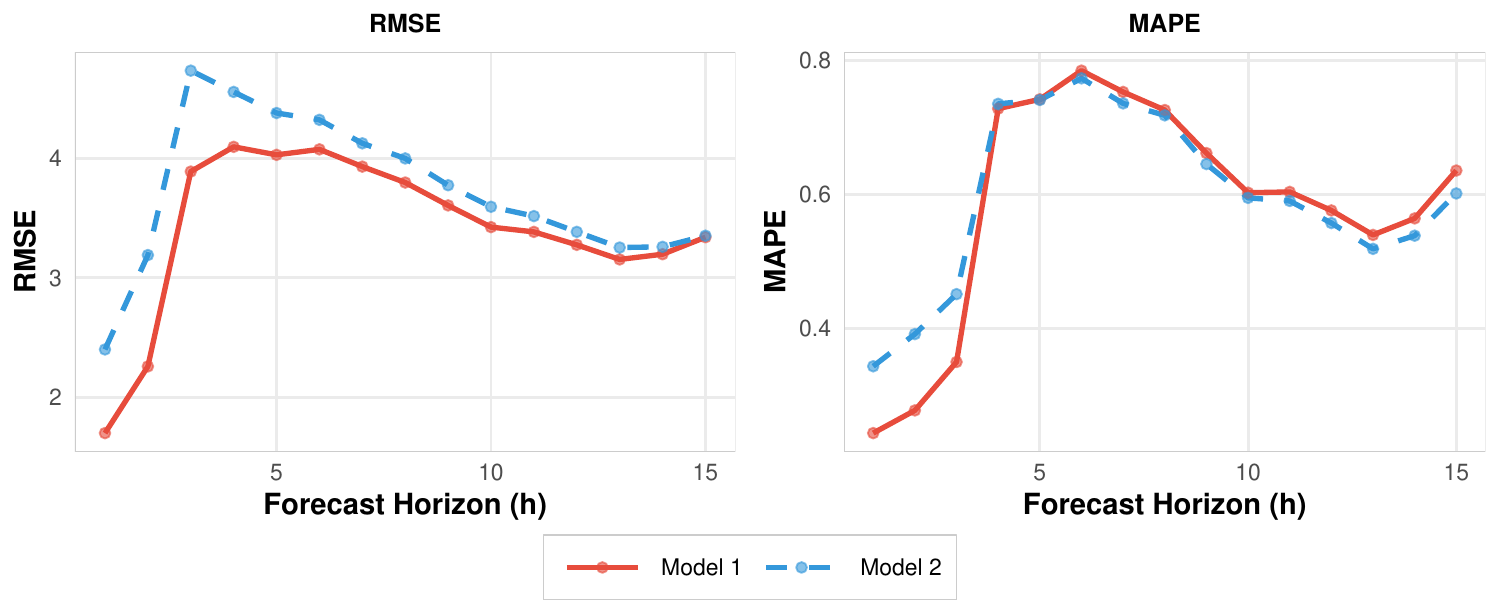}
\caption{Out-of-sample forecast accuracy measure as a function of the horizon for Model 1 and 2.} \label{fig:outofsample}
\end{figure}

\begin{figure}[ht!]
\centering
\includegraphics[width=0.8\textwidth]{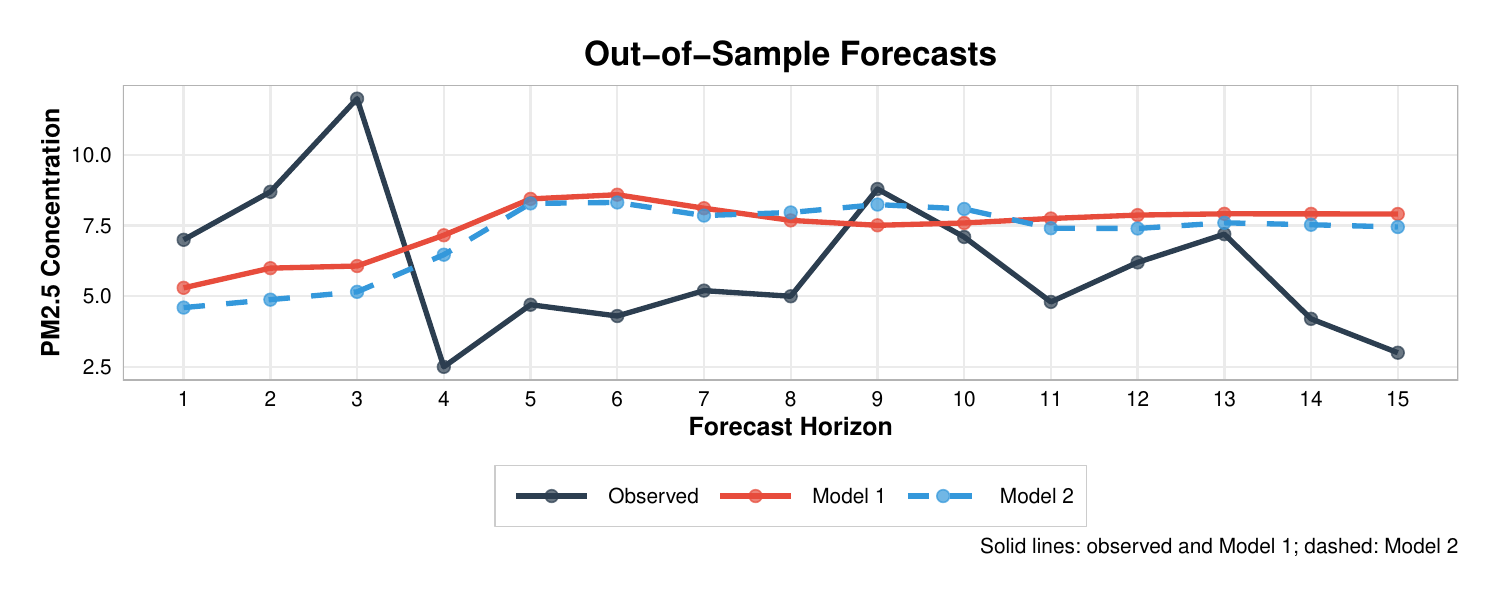}
\caption{Out-of-sample forecast accuracy measure as a function of the horizon for Model 1 and 2.} \label{fig:outofsample_G}
\end{figure}

In this section we evaluate out-of-sample forecasting capabilities of the fitted models.
More explicitly, for each fitted model, $h$-step-ahead predictions $\hat{Y}_{n+h}$, with $h \in \{1,\cdots,15\}$, are computed using only information from the observed sample. The reserved sample is used exclusively to calculate the accuracy measures. Cumulative forecast accuracy up to horizon $h$ is assessed using the RMSE, MAPE and MDA, defined as follows for completeness. Let $\{Y_{n+h}\}_{h = 1}^{15}$ and $\{\hat Y_{n+h}\}_{h = 1}^{15}$ denote the observed (reserved) and predicted values, respectively, then we define
\begin{align*}
\text{RMSE}(h) = \Bigg[\frac{1}{h} \sum_{k=1}^{h} (Y_{n+k} - \hat{Y}_{n+k})^2\Bigg]^{\frac12}, \quad \text{MAPE}(h) = \frac{1}{h} \sum_{k=1}^{h} \biggl|\frac{Y_{n+k} - \hat{Y}_{n+k}}{Y_{n+k}}\biggr|,\\
\text{and}\quad \text{MDA}(h) = \frac{1}{h} \sum_{k=1}^{h} I\bigl(\operatorname{sign}(Y_{n+k} - Y_{n}) = \operatorname{sign}(\hat Y_{n+k} - Y_{n})\bigr), \quad h \in \{1,\dots,15\}.
\end{align*}
Figure~\ref{fig:outofsample} presents the out-of-sample forecast accuracy measures as a function of the forecast horizon for both fitted models whereas Figure~\ref{fig:outofsample_G} present the forecasted values against the observed ones. For RMSE, the proposed Gamma-ARTFIMA model consistently outperforms Model 2 across all horizons. This uniform advantage suggests that the tempering parameter contributes to a slight but systematic improvement in point forecast accuracy, particularly in terms of error magnitude. For MAPE, the performance of both models is remarkably similar, with the Gamma-ARFIMA showing a marginal advantage at certain horizons. However, the differences are minimal, indicating that both models produce relative forecast errors of comparable magnitude. Regarding directional accuracy, the MDA was identical for both models, with a minimum of 0.75 across all horizons. 
The identical MDA performance suggests that both models capture the directional movements of the series equally well. Overall, while both models perform adequately in terms of out-of-sample prediction, the Gamma-ARTFIMA offers a slight yet consistent edge in RMSE, reinforcing the benefits of the tempering parameter for multi-step ahead forecasting.

\FloatBarrier
\section{Application to Amazon stock}\label{amzn}

In this section, we apply the proposed model to the adjusted closing prices ($C_t$) of Amazon (AMZN) stock, from Feburary 18, 2019 to June 30, 2022, yielding $n = 848$ observations. The data are freely available from Yahoo Finance (\url{https://finance.yahoo.com/}) and can be easily retrieved using the \texttt{quantmod} package \citep{quantmod}.

\begin{figure}[ht!]
\centering
\includegraphics[width=0.8\textwidth]{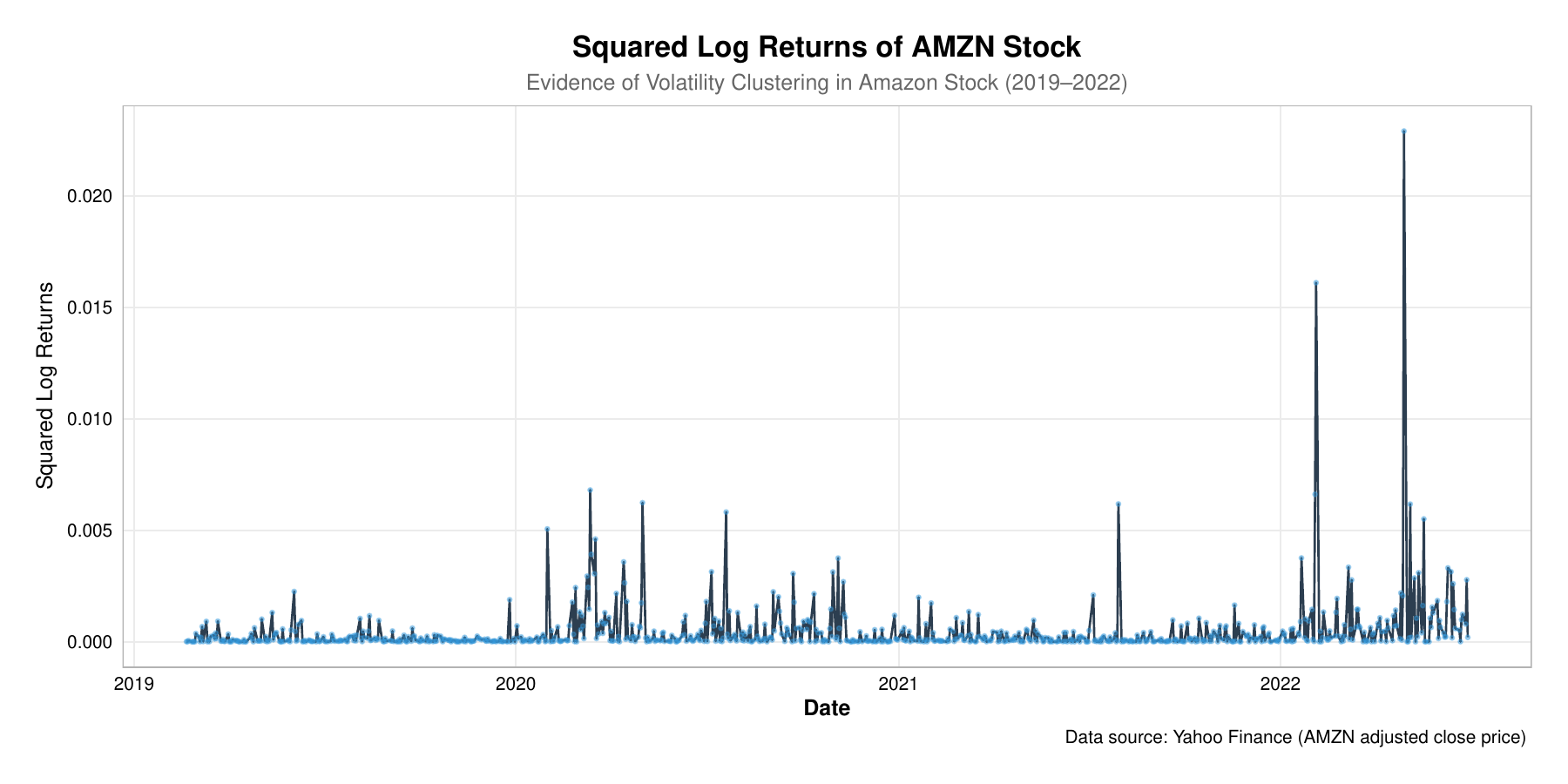}
\includegraphics[width=0.7\textwidth]{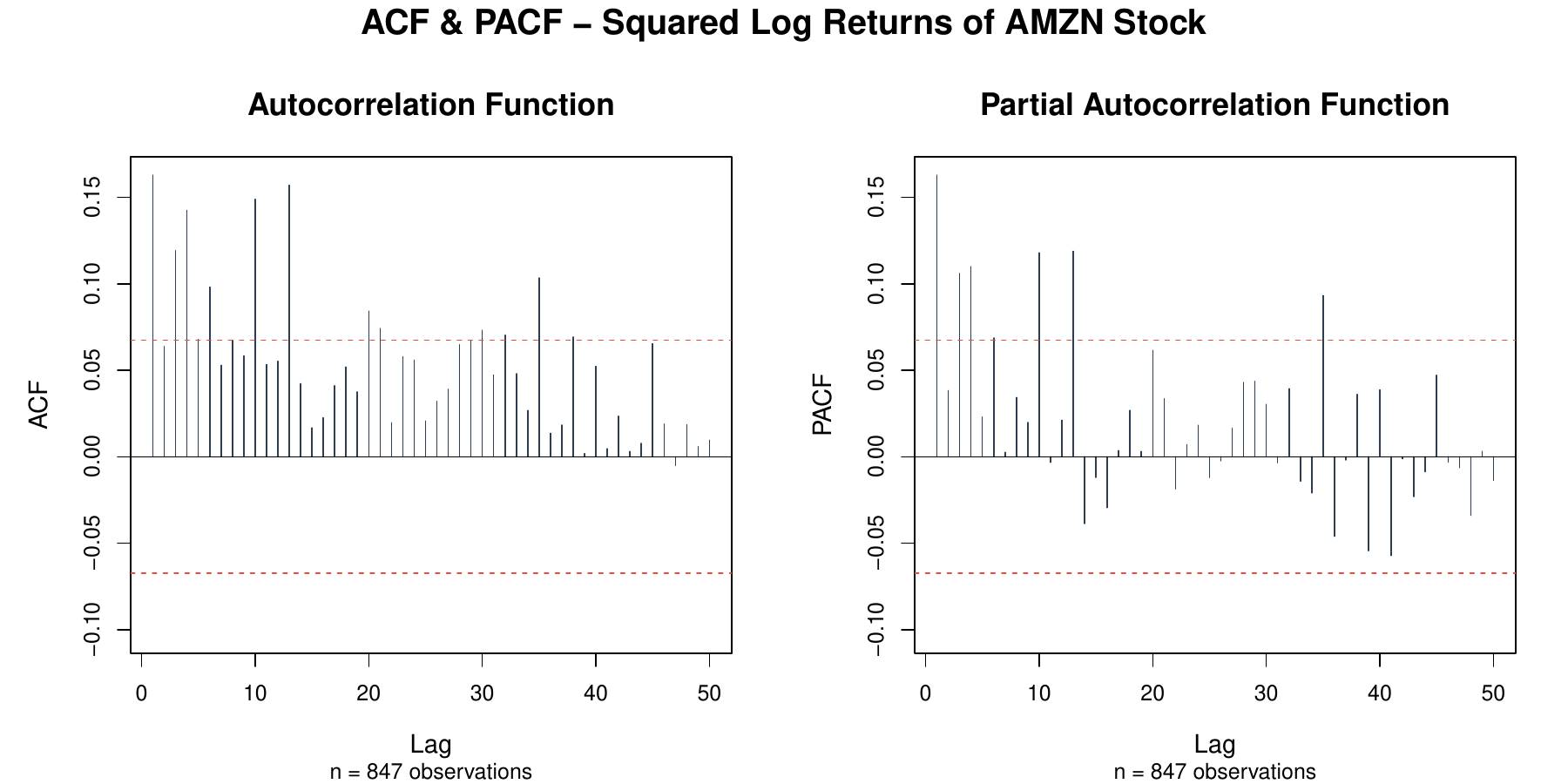}
\caption{Time series plots (top) and ACF and PACF plots (bottom).} \label{basicplots1}
\end{figure}

We consider the squared log-returns defined as $R_t^2 = [\log(C_t / C_{t-1})]^2$, with the first observation lost in the transformation. Figure~\ref{basicplots1} displays the time series of squared log-returns along with their sample autocorrelation (ACF) and partial autocorrelation (PACF) functions.

Squared financial returns are typically characterized by several stylized facts, in particular that (i) they exhibit positive autocorrelation and (ii) the autocorrelation decay slowly, often suggesting long memory. From the plots, and considering the asymptotic confidence intervals, the correlation decay appears too rapid to be characteristic of long-range dependence, despite the aforementioned stylized facts. This observation was already reported in \cite{artfima}, albeit the authors used a Gaussian model for the strictly positive time series. We will investigate this point in greater detail using the proposed Gamma-ARTFIMA model, which appropriately accounts for the strictly positive nature of the data. Additionally, the ACF plots reveal pronounced spikes at lags 10 and 13. Accordingly, we proceed with a backward elimination strategy, as described in Section~\ref{appl}, taking lag 13 as the starting point. Hence, we start by considering the full model with $p = 13$ and $q = 13$, eliminating at each round the single AR or MA coefficient with the highest $p$-value exceeding 0.05. We proceed with the elimination until all remaining coefficients are significant at the 5\% level. Parameters $\lambda$ and $d$ were kept until the very last model, while $\nu$ and $\alpha$ were the only significant coefficients throughout. We also consider a Gamma-ARFIMA model for comparison, applying the same backward elimination procedure.

The backward procedure yielded a Gamma-ARTFIMA$(2,d,\lambda,2)$ model with the restriction $\theta_1 =0 $ and $\phi_2 = 0$. This specification is denoted as ``Model 1'' throughout the remainder of this section. The same procedure yielded a Gamma-ARFIMA$(2,d,2)$, with no restrictions, referred to as ``Model 2''. Interestingly, both models present the same number of parameters.

Parameter estimates and their corresponding $p$-values are reported in Table~\ref{tab:est1}. For Model 1, the estimated tempering parameter $\hat{\lambda} = 0.0525$ is statistically significant, indicating that the persistence in the series is well captured by the tempering mechanism. In addition, the estimated fractional differencing parameter $\hat{d} = 0.8715$ suggests substantial persistence in the transformed series. The joint presence of a positive tempering parameter and a relatively large fractional parameter is consistent with a semi-long memory process, where dependence remains pronounced over short and intermediate horizons but gradually weakens at larger lags \citep{artfima}. The significant coefficients $\phi_1$ and $\theta_2$ further demonstrate that both autoregressive and moving-average effects contribute to the observed dynamics. Overall, Model 1 achieves a slightly lower AIC value ($-12284$ vs $-12216$) and a higher log-likelihood ($6148.1$ vs $6115$), both favoring Model 1.

\begin{table}[ht]
\centering
\caption{Estimated parameters along with respective $p-$values for the fitted GARTFIMA (Model 1), and GARFIMA models (Model 2). Also presented are the AIC and log-likelihood values.}\label{tab:est1}
\vskip.2cm
\begin{tabular}{c|r|r||r|r}
\multirow{2}{*}{Parameter} &\multicolumn{2}{c||}{Model 1} &\multicolumn{2}{c}{Model 2}  \\
\cline{2-5}
& Estimate & $p$-value &  Estimate & $p$-value  \\
 \hline
$\alpha$   &  -9.9333  &  $<0.0001$  &  -8.3122  &  $<0.0001$ \\
$\phi_1$   &  -0.7751  &  $<0.0001$  &   0.4142  &  $<0.0001$ \\
$\phi_2$   &   ---     &     ---     &  -0.7758  &  $<0.0001$ \\
$\theta_1$ &   ---     &     ---     &  -0.4510  &  $<0.0001$ \\
$\theta_2$ & -0.6654   &  $<0.0001$  &   0.7488  &  $<0.0001$ \\
$d$        &  0.8715   &  $<0.0001$  &   0.1220  &  $<0.0001$ \\ 
$\lambda$  &  0.0525   &  $<0.0001$  &    ---    &    ---     \\
$\nu$      &  0.4164   &  $<0.0001$  &   0.3954  &  $<0.0001$ \\
\hline
AIC        &  \multicolumn{2}{c||}{-12284} & \multicolumn{2}{c}{-12216} \\
log-like   & \multicolumn{2}{c||}{6148.1} &  \multicolumn{2}{c}{6115} \\
\hline
\end{tabular}
\end{table}
The Gamma-ARFIMA model (Model 2) yields a considerably smaller estimate of the fractional differencing parameter, $\hat{d} = 0.1220$, implying a weak degree of persistence, despite still pointing to the presence of long-range dependence.

Comparatively, we notice that the estimated values of $\theta_2$ opposite in sign across both models, but their magnitudes differ substantially. The estimates of $\nu$ are consistent across models ($0.4164$ in Model 1 versus $0.3954 $ in Model 2), as are the estimates of $\alpha$, which share the same sign and comparable magnitudes. Both models systematic components are free from unit roots.

\begin{figure}[ht!]
\centering
\includegraphics[width=0.7\textwidth]{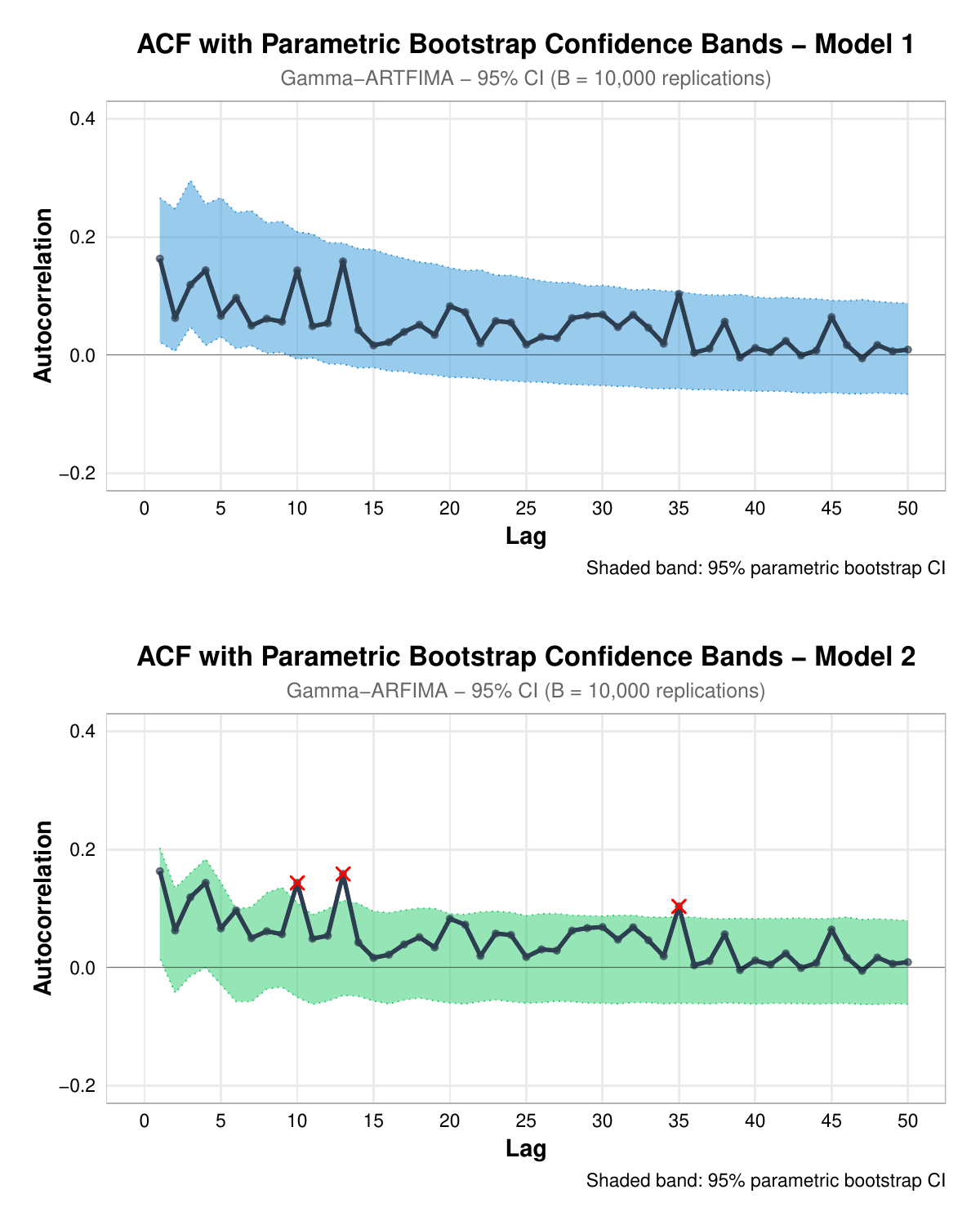}
\caption{Bootstrap confidence bands for the ACF of (top) Model 1 (Gamma-ARTFIMA) and (bottom) Model 2 (Gamma-ARFIMA) for squared log-returns ($R_t^2$) along with the observed time series ACF. Points lying outside the confidence bands are highlighted in red.} \label{confbands2}
\end{figure}

To assess the ability of each model to reproduce the dependence structure of the observed time series, we apply a parametric bootstrap approach to obtain 95\% confidence bands for the ACF of each fitted model. The procedure is conducted under the same settings as in Section~\ref{appl}, except that we consider $n = 750$. The resulting confidence bands, along with the empirical ACF of the observed series, are displayed in Figure~\ref{confbands2}. Autocorrelations falling outside the bands are highlighted in red.

From the Figure~\ref{confbands2}, we observe that Model 1 provides an excellent representation of the autocorrelation structure, as all empirical autocorrelations from lags 1 to 50 lie within its confidence bands. Although Model 2 also provide a reaonably good representation, we observe the empirical ACF exceeding the confidence limits at lags 10, 13, and 35. Consequently, despite Model 2 including additional ARMA coefficients, the extended flexibility provided by the tempering parameter in Model 1 allow the model to provide a better fit for the the observed autocorrelation pattern.

\FloatBarrier
\subsection{Residual Analysis}

To formally evaluate whether any residual dependence remains, we examined the martingale difference property of the residuals using  the Dominguez-Lobato martingale difference test \citep{DL}, considering both the Cram\'er-von Mises (CM) and Kolmogorov-Smirnov (KS) test statistics, as implemented in the \texttt{BTSR} package. For each test, the corresponding p-values were obtained using 5,000 bootstrap replications. The results are summarized in Table~\ref{tab:mart2}. Across all considered tests, the null hypothesis of a martingale difference sequence cannot be rejected, indicating that the residuals from both models exhibit no significant remaining serial dependence. These findings suggest that the conditional mean dynamics are effectively captured by the proposed model specifications and provide further evidence supporting the adequacy of both fitted models.\\

\begin{table}[ht]
\centering
\caption{$p$-values of the martingale difference tests applied to the simple residuals of the Gamma-ARTFIMA and Gamma-ARFIMA models for Squared Log Returns of AMZN Stock with based on 5,000 bootstrap iterations.}\label{tab:mart2}
\vskip.2cm
\begin{tabular}{c|cc}
\hline
 & Cp & Kp     \\
\hline
Model 1 & 0.3822 & 0.3852   \\
Model 2 & 0.0616 & 0.0794   \\
\hline
\end{tabular}
\end{table}

\subsection{In-Sample Forecast}

In terms of in-sample forecasting, Model 1 performs slightly better, with an RMSE of 0.0008 vs. 0.0011 and a MAPE of 0.4845 compared to 1.4432. These results indicate that the Gamma-ARTFIMA model provides more accurate in-sample forecasts overall, consistent with the information criteria, residual diagnostics, and autocorrelation analysis, all of which offer stronger support for the Gamma-ARTFIMA specification. 

\subsection{Out-of-Sample Forecast}
\begin{figure}[ht!]
\centering
\includegraphics[width=0.8\textwidth]{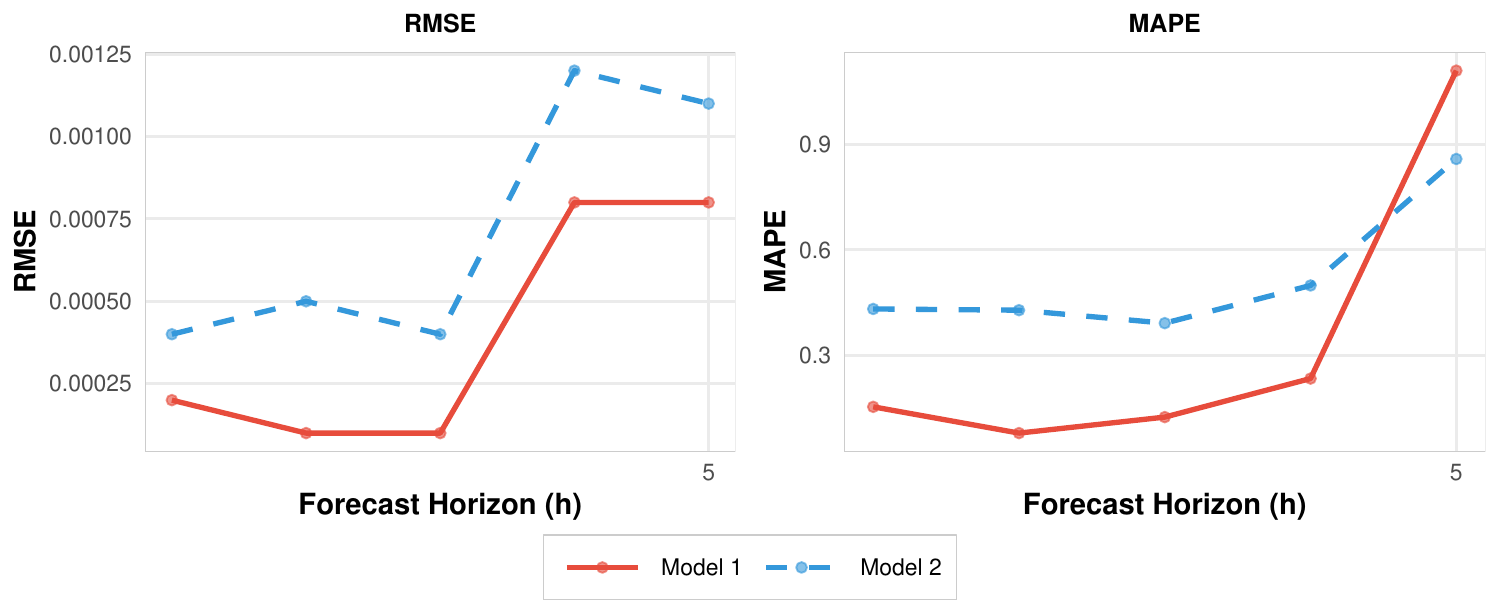}
\caption{Out-of-sample forecast accuracy measures as a function of the horizon for Model 1 and 2.} \label{fig:outofsample1}
\end{figure}

We proceed with out-sample forecast, considering a short term horizon of 5 steps-ahead. Figure~\ref{fig:outofsample1} presents the cumulative out-of-sample forecast accuracy measures up to horizon (h) for the two fitted models. The forecast performance is evaluated using the RMSE and MAPE, computed cumulatively over the prediction horizons. 

The proposed Gamma-ARTFIMA model exhibits uniformly lower RMSE values than the Gamma-ARFIMA model across all forecast horizons, indicating superior predictive accuracy in terms of error magnitude. For MAPE, the proposed Gamma-ARTFIMA model consistently outperforms the Gamma-ARFIMA model across, up to lag 5. At $h=5$ the cumulative MAPE of Model 2 surpasses Model 1's. 

Overall, in terms of out-of-sample forecasting accuracy, there is a tangible advantage for Gamma-ARTFIMA model throughout the entire prediction horizon.

\begin{figure}[ht!]
\centering
\includegraphics[width=0.8\textwidth]{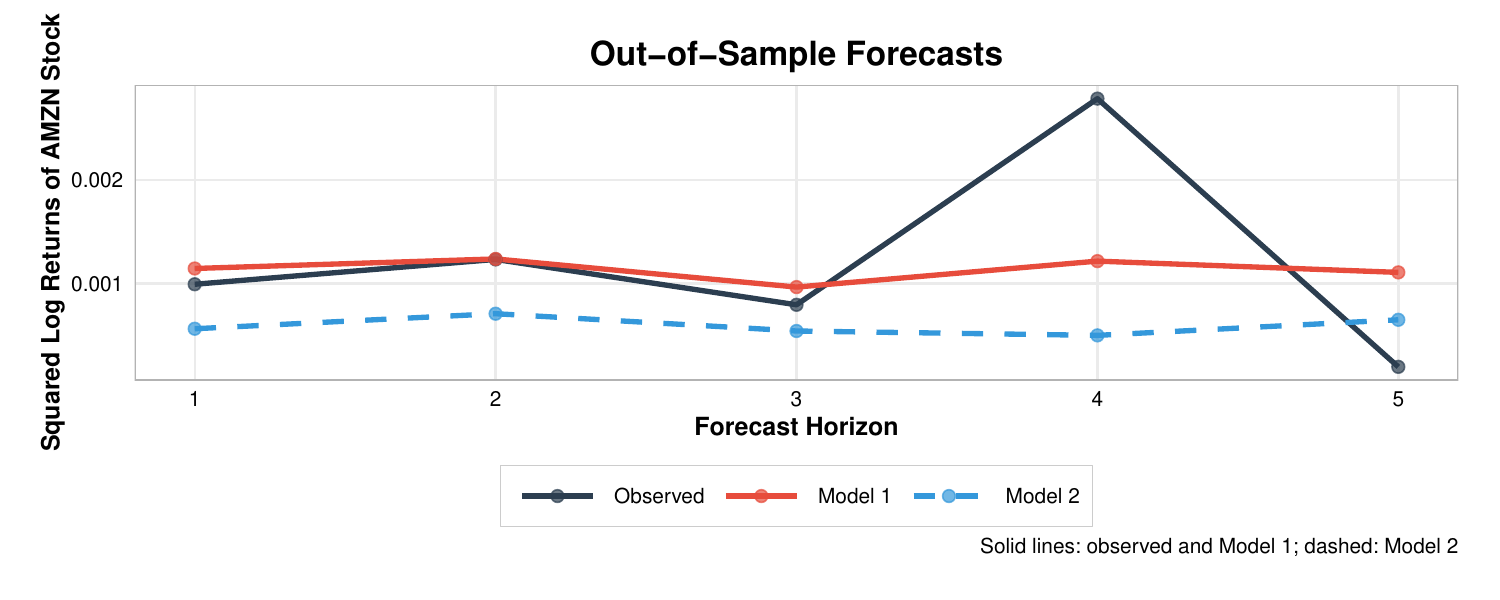}
\caption{Out-of-sample forecast accuracy measure as a function of the horizon for Model 1 and 2.} \label{fig:outofsample2}
\end{figure}

\section{Conclusions}\label{sec:end}

This paper introduced a novel class of observation-driven models -- the GARTFIMA models -- that extends the GARMA and GARFIMA frameworks by incorporating a tempered fractional differencing operator into the systematic component. This generalization enables the modeling of non-Gaussian time series that exhibit semi-long memory behavior, effectively capturing strong short-term persistence while allowing for faster decay at longer lags through the tempering parameter $\lambda$. The proposed class accommodates a wide range of data types, including positive, double-bounded, and count-valued outcomes, while preserving the flexibility and interpretability of the GARMA framework.

Parameter estimation was developed via partial maximum likelihood, with closed-form expressions for the score vector and conditional information matrix. A Monte Carlo simulation study demonstrated that the PMLE performs well in finite samples, with estimates stabilizing as the sample size increases. The empirical applications to PM$_{2.5}$ pollution data and Amazon stock squared log-returns illustrated the practical utility of the proposed model. In both cases, the GARTFIMA model outperformed its GARFIMA counterpart, yielding lower AIC values, superior residual diagnostics, and more accurate out-of-sample forecasts. The bootstrap-based ACF analysis further revealed that the proposed model faithfully reproduces the dependence structure of the observed series, whereas the GARFIMA model systematically fails to capture key autocorrelation patterns. Overall, the GARTFIMA class offers a coherent and flexible framework for analyzing non-Gaussian time series with tempered long-range dependence, bridging the gap between generalized linear modeling and fractional tempering.

\subsection*{Acknowledgments}
G. Pumi and T.S. Prass gratefully acknowledge the financial support received by the Conselho Nacional de Desenvolvimento Cient\'ifico e Tecnol\'ogico -- CNPq Brasil  -- Bolsa de Produtividade em Pesquisa - Proc. 303281/2025-1 (Pumi) and 305886/2025-8 (Prass). 

\bibliographystyle{apalike}
\bibliography{bib}
\appendix
\section{Calculations for the Gamma-ARTFIMA model}
In this section we exemplify the calculations related to the distribution specific term $c(y_t;\nu)$ considering the gamma distribution. Reparameterization of the gamma in terms of its mean, facilitates the calculations. in this case, the density becomes
\begin{equation}\label{dgamma}
f(y;\mu_t,\nu)= \frac{1}{\Gamma(\frac1\nu)} \left(\frac{1}{\nu\mu_t}\right)^{\frac1\nu} y^{\frac1{\nu}-1} \exp\bigg\{-\frac{y}{\nu\mu_t}\bigg\} I(y>0), 
\end{equation}
that is, the shape and scale parameters are $\frac1\nu$ and $\nu\mu_t$. From \eqref{dgamma} we have that
\begin{equation*}
c(Y_{t},\nu)= \left(\frac{1}{\nu}-1\right)\log(Y_t) -\frac{\log(\nu)}{\nu} -\log\big(\Gamma(1/\nu)\big),
\end{equation*}
from which it is straightforward to show that 
\begin{equation*}
\frac{\partial c(Y_t,\nu)}{\partial \nu}
= \frac{1}{\nu^{2}}
\big[
\log(\nu)-\log(Y_t) - 1 + \psi(1/\nu)
\big],
\end{equation*}
and
\begin{equation}\label{secdc}
\frac{\partial^2 c(Y_t,\nu)}{\partial\nu^2} = \frac{ 2\log(Y_t) -2\log(\nu) +3 -2\psi(1/\nu) }{\nu^3} - \frac{\psi'(1/\nu)}{\nu^4},
\end{equation}
where $\psi(z):=\frac{d}{dz}\log\big(\Gamma(z)\big)$ is the digamma function and $\psi'(z)=\frac{d}{dz}\psi(z)$ is the tri-gamma function. By taking conditional expectation in \eqref{secdc}, and from standard result for gamma distributed  random variables, we have $\E\big(\log(Y_t)|\F_{t-1}\big) = \psi(1/\nu) + \log(\nu\mu_t)$, so that
\begin{equation*}
\E\bigg( \frac{\partial^2 c(Y_t,\nu)}{\partial\nu^2} \bigg| \F_{t-1} \bigg) = \frac{ 2\log(\mu_t)+3 }{\nu^3} - \frac{\psi'(1/\nu)}{\nu^4}.
\end{equation*}
Furthermore, from \eqref{dgamma}, we conclude that $h(x)=-\frac1x$ and $b(x)=-\log(-x)$. Consequently, \eqref{nu} simplifies to
\begin{align*}
\E\bigg(\frac{\partial^2 \ell_t(\bs\gamma)}{\partial \nu^2}\bigg|\F_{t-1}\bigg)
&=-\frac{2Y_t}{\mu_t\nu^3}- \frac{\psi'(1/\nu)}{\nu^4}.
\end{align*}

\end{document}